\documentclass[12pt,letterpaper]{article}
\pdfoutput=1
\usepackage{graphicx,array}
\usepackage{color}
\usepackage{latexsym}
\usepackage{amsthm}
\usepackage{amsmath}
\usepackage{amssymb}
\usepackage{hyperref}
\usepackage{bbold}

\def\simgt{\mathrel{\lower2.5pt\vbox{\lineskip=0pt\baselineskip=0pt
           \hbox{$>$}\hbox{$\sim$}}}}
\def\simlt{\mathrel{\lower2.5pt\vbox{\lineskip=0pt\baselineskip=0pt
           \hbox{$<$}\hbox{$\sim$}}}}

\newcommand{\be}{\begin{equation}}
\newcommand{\ee}{\end{equation}}
\newcommand{\bea}{\begin{eqnarray}}
\newcommand{\eea}{\end{eqnarray}}

\hypersetup{colorlinks,citecolor= blue,linkcolor= black}

\begin{document}

\begin{center}
{\LARGE\bf The Weak Scale from BBN
}
\bigskip\vspace{1cm}{

\large Lawrence J. Hall$^{1,2}$, David Pinner$^{1,2,3}$, and Joshua T. Ruderman$^{1,2,4}$}
 \\[7mm]
 {\it 

 $^1$Berkeley Center for Theoretical Physics, Department of Physics, \\
     $^2$Theoretical Physics Group, Lawrence Berkeley National Laboratory, \\ University of California, Berkeley, CA 94720, USA \\
     $^3$Princeton Center for Theoretical Science, \\ Princeton University, Princeton, NJ 08544, USA \\
     $^4$Center for Cosmology and Particle Physics, \\ Department of Physics, New York University, New York, NY 10003 \\
     } 
   \end{center}

\bigskip \bigskip
\centerline{\large\bf Abstract}
\bigskip   
The measured values of the weak scale, $v$, and the first generation masses, $m_{u,d,e}$, are simultaneously explained in the multiverse, with all these parameters scanning independently.  At the same time, several remarkable coincidences are understood.  Small variations in these parameters away from their measured values lead to the instability of hydrogen, the instability of heavy nuclei, and either a hydrogen or a helium dominated universe from Big Bang Nucleosynthesis.  In the 4d parameter space of $(m_u,m_d,m_e,v)$, catastrophic boundaries are reached by separately increasing each parameter above its measured value by a factor of $(1.4,1.3,2.5,\sim5)$, respectively.
The fine-tuning problem of the weak scale in the Standard Model is solved: as $v$ is increased beyond the observed value, it is impossible to maintain a significant cosmological hydrogen abundance for any values of $m_{u,d,e}$ that yield both hydrogen and heavy nuclei stability. 

For very large values of $v$ a new regime is entered where weak interactions freeze out before the QCD phase transition. The helium abundance becomes independent of $v$ and is determined by the cosmic baryon and lepton asymmetries.  To maintain our explanation of $v$ from the anthropic cost of helium dominance then requires universes with such large $v$ to be rare in the multiverse. Implications of this are explored, including the possibility that new physics below 10 TeV cuts off the fine-tuning in $v$.

\begin{quote} \small

\end{quote}

\newpage

\newpage

\tableofcontents
\newpage
%%%%%%%%%%%%%%%%%%%%%%%%%%%
\section{Introduction}
\label{sec:intro}
%%%%%%%%%%%%%%%%%%%%%%%%%%%%

All results from particle physics experiments and cosmological observations can be described consistently by an effective theory of the Standard Model and General Relativity, (SM+GR). Furthermore, this effective theory may be valid up to a very high mass scale $\Lambda_{SM}$, that exceeds the weak scale $v$ by many orders of magnitude; the physics of inflation, baryogenesis and dark matter may be associated with mass scales even larger than $\Lambda_{SM}$.   While consistent with all data, such an effective theory has fine-tuning of at least $v^2/\Lambda_{SM}^2$ in the weak scale, and $\Lambda_{CC}/\Lambda_{SM}^4$ in the cosmological constant $\Lambda_{CC}$.   The prevailing view of the last quarter of the 20th century was that $\Lambda_{SM}$ would turn out to be of order the weak scale and that a natural theoretical understanding for a very small or zero cosmological constant would emerge.   The discovery of dark energy in 1998 and the absence of a Higgs boson or new physics at LEP by 2000 produced cracks in this conventional view, but still the expectation was that the LHC was likely to uncover a natural understanding for the weak scale.  

The discovery of a perturbatively coupled Higgs boson at the LHC, and the absence of any signs of breakdown of the SM at LHC or elsewhere, places this prevailing view under further stress.  After run 1 of the LHC, the fine-tuning of $v$ is becoming problematic. 
While a discovery at the next run of the LHC could dramatically change our understanding of the origin of $v$, in the absence of any discovery of physics beyond the SM, and of any natural explanation for the amount of dark energy, the development of a theoretical framework to understand the fine-tunings of both $v$ and $\Lambda_{CC}$ is highly motivated.  

The multiverse, based on eternal inflation and the string landscape, may provide such a framework.  If observers are rare in the multiverse, then those universes that do have observers can contain parameters that appear to be finely tuned. In particular, it has been argued that most universes do not contain large scale structure, so that observed values of $\Lambda_{CC}$ are tuned~\cite{Weinberg:1987dv}, and most universes do not contain complex nuclei, so that $v$ is observed to be fine-tuned~\cite{Agrawal:1997gf}.   If our observed value of $\Lambda_{CC}$ is increased by about 2 orders of magnitude galaxies fail to form~\cite{Martel:1997vi} and if $v$ is increased by about 50\% there are no bound complex nuclei~\cite{Damour:2007uv}.   So far the fine-tuning problems of the cosmological constant and weak scale have resisted solutions by means of symmetries; their persistence provides evidence for the multiverse.

These key results for $\Lambda_{CC}$ and $v$ were each obtained by studying multiverses where only a single parameter scans.   In (SM+GR) there are only three mass parameters $\Lambda_{CC}, v$ and the Planck mass $M_{\text{pl}}$, so these results would also follow in (SM+GR) in a landscape that only allows dimensional parameters to scan~\cite{ArkaniHamed:2005yv}.   However, in more general landscapes dimensionless parameters scan.  Furthermore, in theories that go beyond (SM+GR) small dimensionless parameters, such as Yukawa couplings~\cite{Froggatt:1978nt, ArkaniHamed:1999dc} and the primordial density perturbations, are understood in terms of ratios of disparate mass scales, so that they would scan even in the restricted landscapes of~\cite{ArkaniHamed:2005yv}.  

\begin{figure}[h!]
\begin{center} \includegraphics[width=1.00 \textwidth]{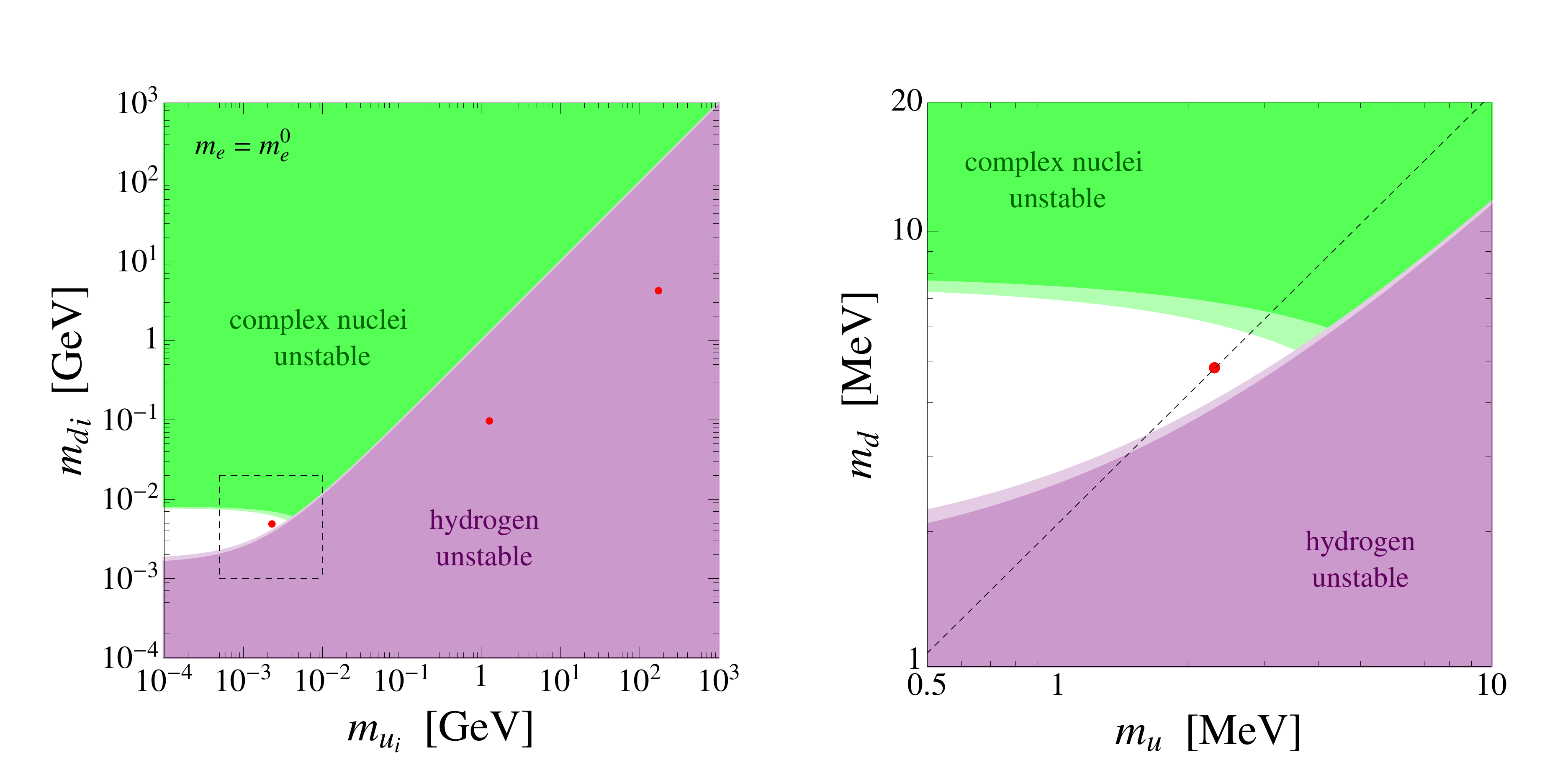}\end{center}
\caption{ \label{fig:nuclearboundaries}
The anthropic requirements on $(m_u, m_d)$ for the stability of complex nuclei  (upper, green) and the stability of hydrogen (lower, purple), with other parameters held fixed.  In the left panel the three red dots correspond to the observed masses of the three quark generations.  The light shading gives the $1 \sigma$ theoretical uncertainties in each boundary, which, of course, only apply to the lightest generation.  The region in the dashed square is shown expanded in the right panel, where the dashed line corresponds to varying $v$ with Yukawa couplings held fixed. }
\end{figure}

The observed values of the up and down quark masses place our universe on the edge of {\it both} hydrogen stability and complex nuclei stability, as shown in Figure~\ref{fig:nuclearboundaries}, suggesting {\it independent} scanning of these masses.   The observed values of $(m_u, m_d)$ lie near the tip of a cone formed by these boundaries.   Given that the SM Yukawa couplings vary over 5-6 orders of magnitude, it is remarkable that  variations of only $\sim 30$\% lead to a catastrophic change of atomic physics.  If flavor results purely from symmetries, this closeness to the tip of the cone is accidental -- the lowest red dot would be expected to lie far from the tip of the cone.  If it were in the shaded region of Figure~\ref{fig:nuclearboundaries}, the universe would not contain observers; if it were in the unshaded region, observers would not find that they lived close to catastrophic boundaries.  However, in a multiverse with up and down Yukawa couplings scanning, with distributions favoring a large up coupling, the most probable observed universes lie near the tip of the cone.  Note that if only $v$ scans, corresponding to moving alone the dashed line of the right panel, our proximity to the complex nuclei boundary can be understood, but the proximity of the line to the tip of the cone is accidental.  Two scanning parameters, such as $(m_u,m_d)$, can naturally explain our proximity to the tip of the cone, while one scanning parameter cannot.

A key question is whether the multiverse understanding of the finely-tuned values of $\Lambda_{CC}$ and $v$ is robust to scanning many parameters.    In the case of $\Lambda_{CC}$ the argument appears to fail if the energy density at virialization, $\rho_{vir}$, is allowed to scan.  The large scale structure catastrophic boundary determines $\Lambda_{CC}/\rho_{vir}$, so there is a runaway direction to large values of $\rho_{vir}$ and $\Lambda_{CC}$.  One way to avoid this conclusion is to find another catastrophic boundary so that both $\Lambda_{CC}$ and $\rho_{vir}$ are both determined~\cite{Tegmark:2005dy}.\footnote{Another possibility is that $\Lambda_{CC}$ is determined by a different environmental argument, which seems preferred since we observe $\Lambda_{CC}$ about two orders of magnitude away from the large scale structure boundary. The latter is accomplished by using the causal patch measure, giving a statistical prediction that $\Lambda_{CC}$ dominates the energy density of the universe in the era containing most observers, solving the ``Why Now?" problem~\cite{Bousso:2007kq}.} 

Similarly, in the case of the electroweak scale the argument from complex nuclei fails if the up and down quark Yukawa couplings, $y_{u,d}$, scan.  In this case nuclear physics environmentally selects $m_{u,d} = y_{u,d} \, v$, as shown in Figure~\ref{fig:nuclearboundaries}, leaving open a possible runaway direction to large $v$ and small $y_{u,d}$.  Thus, the crucial question is whether there are other environmental effects that depend on different combinations of $y_{u,d}$ and $v$.  While we know of no environmental boundaries that select for $y_{u,d}$, there are three well-known astrophysical phenomena involving the amplitude for weak interaction processes, $G_F \sim 1/v^2$, that could be anthropically relevant~\cite{Carr:1979sg}.

If $G_F$ were too small (large) then neutrinos produced in the final stages of stellar collapse might easily escape (get trapped) preventing supernova explosions which allow heavy elements produced in early generations of stars to be recycled into later planetary systems.  The role of neutrinos in supernova explosions has been extensively studied, and it is still not known whether they play a critical role (see for example the review~\cite{Janka:2012wk}), so we do not pursue this possibility.   As $G_F$ varies so does the rate for the $pp \rightarrow d e^+ \nu$ reaction that initiates the pp cycle for energy production in main sequence stars.  However, long-lived stars can still be fueled by the pp chain: the central temperature of the star can adjust to compensate for the change in the weak amplitude without substantially altering the luminosity.  As the weak scale is increased from the observed value, there is no nearby anthropic boundary from pp stellar burning.

\begin{figure}[h!]
\begin{center} \includegraphics[width=0.7 \textwidth]{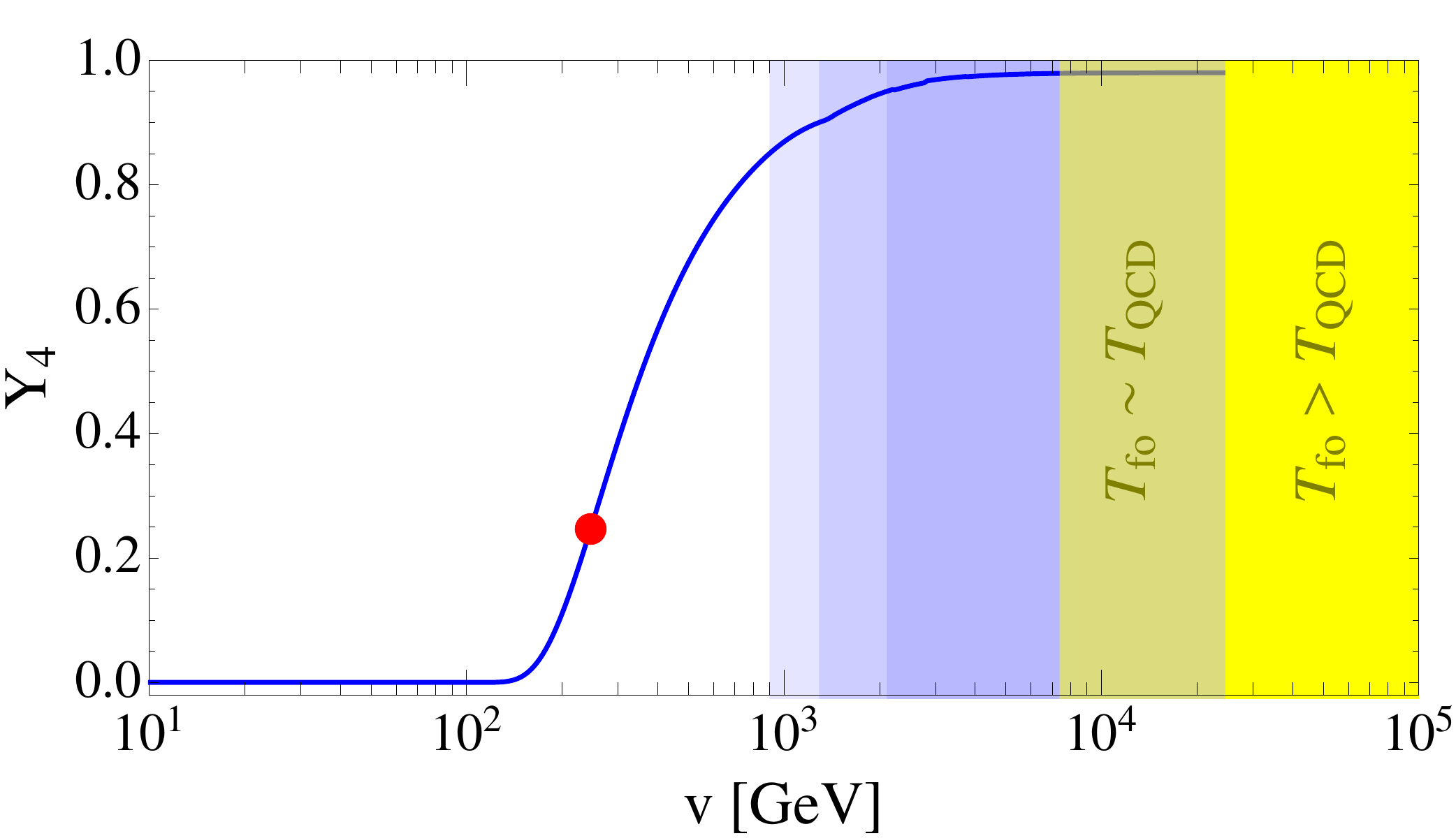}\end{center}
\caption{ \label{fig:BBNv}
The fraction of baryonic mass processed to $^4$He during BBN as a function of the weak scale, with other physically relevant parameters, including $m_{u,d,e}, \Lambda_{\text{QCD}}$ and $M_{\text{pl}}$, held fixed. The red dot shows our universe, and appears on the steeply rising portion of the curve.  The blue shading shows regions with more than $(85, 90, 95)$\% of baryons in helium.  The yellow region has freeze-out for $u \leftrightarrow d$ conversion occurring before the QCD phase transition while, in the region of overlapping blue/yellow shading, the ordering of the freeze-out and phase transition is uncertain. 
}
\end{figure}

In our universe 25\% of baryons are processed into helium during the era of Big Bang Nucleosynthesis (BBN).  The helium fraction is highly sensitive to scanning the weak scale $v$ around our observed value, as shown in Figure~\ref{fig:BBNv}.  In particular, as $v$ is increased the helium fraction rapidly rises above 90\%.  While it is unclear whether even a 100\% helium universe is catastrophic, a large increase in the helium fraction does have consequences that clearly tend to suppress observers: halo cooling becomes slower, long-lived hydrogen burning stars become rarer, and hydrogen as a building block of life becomes rarer.   The fine-tuning of the weak scale in the SM implies that if $v$ scans most universes will have a large value of $v$, giving a probability force to the right in Figure~\ref{fig:BBNv}.   Remarkably, {\it our universe lies on the steep part of the curve in Figure~\ref{fig:BBNv}}, where universes transition from dominantly hydrogen to dominantly helium, suggesting an anthropic cost of excessive helium.   In this paper we explore the possibility that the weak scale originates from environmental selection at BBN\@.  We are interested in modest variations in $v$ and assume that huge variations, as in the weakless universe~\cite{Harnik:2006vj}, give universes less probable than our own.

The transition from a dominant hydrogen to a dominant helium universe occurs when the freeze-out temperature of neutron-proton interconversion is close to the neutron-proton mass difference, yielding the parametric prediction
\bea
v \sim (m_n - m_p)^{3/4} \; M_{\text{pl}}^{1/4}.
\label{eq:v}
\eea
This relation is of course a correlation among $(v, m_n - m_p, M_{\text{pl}})$.  However, $m_n - m_p$ and $M_{\text{pl}}$ are likely selected via phenomena that do not involve the weak interactions, so that it is the BBN helium boundary that prevents the runaway of the weak scale to large values.  The weak scale is indeed fine-tuned, but no more than is typically necessary in the multiverse for observers to exist.

Given that $(y_u, y_d)$ and $v$ could vary over {\it many} orders of magnitude, the extreme closeness of the observed values to the catastrophic boundaries in each of Figures~\ref{fig:nuclearboundaries} and~\ref{fig:BBNv} provides evidence for environmental selection in a multiverse.  Our purpose in this paper is to study and assess this evidence. 

In Sections~\ref{sec:distrib} -- \ref{sec:scanetaMP} we restrict our attention to variations of $v$ up to $30 - 100$ times the observed value of the weak scale, $v_0$, so that the relevant weak interaction freezeout occurs {\it after} the QCD phase transition.  We also keep the second and third generation Yukawa couplings fixed so that heavy flavors are not relevant for BBN in this range of $v$. As we vary an increasing set of parameters away from the observed values, we explore the form of the observer boundary, beyond which observers are either absent, or severely constrained.   In the next section we discuss the relevant multiverse distribution functions, and in Section~\ref{sec:atomic} we study selection of $m_{u,d,e}$ at the atomic boundaries.  We then address the BBN helium boundary: 
\begin{itemize}
\item 
With $v/v_0 <  30-100$ the nuclear abundances resulting from BBN depend on the masses $m_{u,d,e}$ as well the weak scale $v$, and in Section~\ref{sec:BBNlowTF} we investigate whether the BBN understanding of $v$ persists when the Yukawa couplings also scan.  The helium boundary becomes a surface in the four dimensional parameter space of $(m_{u,d,e},v)$, and the problem is tractable largely because $m_{u,d,e}$ are already highly constrained by the atomic boundaries.   BBN depends on two other key parameters, the Planck scale $M_{\text{pl}}$ and the baryon asymmetry $\eta$.  In Section~\ref{sec:scanetaMP} we demonstrate possible runaway behavior if these parameters scan, and investigate physical mechanisms that prevent such runaways.
\end{itemize}

In Section~\ref{sec:BBNhighTF} we consider $v/v_0 > 30-100$, so that the weak interaction process relevant for the neutron to proton ratio freezes out {\it before} the QCD phase transition:
\begin{itemize}
\item
With $v/v_0 > 30-100$ the neutron to proton ratio is determined by particle anti-particle asymmetries, and is independent of $m_{u,d,e}$, so that (\ref{eq:v}) no longer holds, as illustrated to the right in Figure~\ref{fig:BBNv}.   We discover that universes with a weak scale $10^2 - 10^3$ times larger than in our universe are typically not dominated by $^4$He.  Nevertheless, we argue that our proximity to the helium boundary, shown by the red dot in Figure~\ref{fig:BBNv}, is still significant for explaining the observed value of $v$, and we discuss a variety of possible reasons why we do not find ourselves in a universe with a much larger value of v.
\end{itemize}
Finally, in Section~\ref{sec:HF} we consider possible effects on the helium abundance of scanning heavy flavor Yukawa couplings.  Conclusions are drawn in 
Section~\ref{sec:concl}, and in the appendix we study the sensitivity of $M_{\text{pl}}/\Lambda_{\text{QCD}}$ to scanning of the quark and lepton masses.

%%%%%%%%%%%%%%%%%%%%%%%%%%%
\section{Scanning Parameters and Multiverse Distributions}
\label{sec:distrib}
%%%%%%%%%%%%%%%%%%%%%%%%%%%%

For a set of scanning parameters $x_i$ the probability distribution in the multiverse is
\be
dP = f(x_i) \, n(x_i) \, d \ln x_i
\label{eq:P}
\ee
where $f$ is the a priori distribution of universes in the multiverse with parameter set $x_i$, $n$ is a weighting proportional to the number of observers in a universe with these parameters, and we assume a sufficient density of vacua to justify the continuous limit.

In the multiverse explanation for the cosmological constant~\cite{Weinberg:1987dv} it is sufficient to first study a landscape where only the cosmological constant scans.  However, for a multiverse explanation of the weak scale via BBN, it is {\it not} sufficient to study a landscape where only the weak scale scans.  If only $v$ scans, then the the observer region is determined by nuclear stability, and the effects of BBN on the weighting factor $n(v)$ are subdominant.  Hence we {\it must} also consider variations in the Yukawa couplings of the first generation quarks, so that the minimal landscape involves the scanning parameters $x_i = (y_{u,d},v)$.  We also allow the electron Yukawa coupling $y_e$ to scan as the electron mass is highly relevant to both BBN and hydrogen stability boundaries.  A modest increase in the electron mass leads to both an increase in helium production and to hydrogen instability.  Hence for most of this paper we take the scanning parameter set to be $x_i = (y_a,v)$, with $a=(u,d,e)$, having a probability distribution
\be
dP \, = \, f(m_a/v,v) \; n_{nuc} (m_a) \; n_{BBN} (Y_4) \; n_{other}(m_a,v) \;\;  d \ln m_a \; d\ln v
\label{eq:P2}
\ee
where we find it convenient to change variables to $(m_a,v)$.   For the weighting $n$, we include factors from nuclear stability, helium production at BBN and possible contributions from other physical effects, such as supernova explosions and stellar burning. Furthermore, $n_{nuc}$ is independent of $v$ and we have assumed that $n_{BBN}$ depends only on the helium abundance, $Y_4(m_a,v)$. While $n_{other}$ could depend on both $v$ and $m_a$, we will set it to unity for most of our analysis.   The hydrogen and complex nuclear stability boundaries are quite sharp, so that we can approximate $n_{nuc}$ to be unity inside the observer region $\cal{O}$ and zero elsewhere.  The region $\cal{O}$ is shown white in Figure~\ref{fig:nuclearboundaries}, where it is projected on to the $(m_u, m_d)$ plane, and is more fully explored in Figure~\ref{fig:ude}. 

How, then, are we to understand Figure~\ref{fig:BBNv}, which shows the helium abundance with $v$ scanning but $m_{u,d,e}$ held fixed?  Since $y_a \propto 1/v$, this corresponds to a special slice through the 4d parameter space.  Clearly, the correct analysis is to study helium production in the landscape of (\ref{eq:P2}), and this is the core of the paper and is presented in Section~\ref{sec:BBNlowTF}.  Nevertheless, Figure~\ref{fig:BBNv} provides a very simple approximation to how the weak scale may result from environmental selection at the helium boundary.  The values of $m_a$ that dominate the distribution (\ref{eq:P2}) may be such that the corresponding values of $Y_4(m_a,v)$ are not far from $Y_4(m_{a0},v)$, where $m_{a0}$ are our observed values of the masses.  In this case one can integrate (\ref{eq:P2}) over $m_a$, ignoring the variation in $Y_4$, obtaining 
\be
dP = f_v(v) \, n_{BBN} (v)  \, d\ln v  \hspace{0.5in} \mbox{with} \hspace{0.5in}
f_v(v) = \left[  \int_{\cal{O}} f(m_a/v,v) \, d \ln m_a \right] .
\label{eq:P3}
\ee
It is in this sense that Figure~\ref{fig:BBNv} is to be understood:
$f_v$ is an effective distribution for $v$, and may be much milder than the quadratic behavior $f(y_a,v) \propto v^2$ expected from fine-tuning.  Hence the probability force $F_v = \partial \ln f_v/ \partial \ln v$, which must be positive at $v=v_0$ explaining why the red dot in Figure~\ref{fig:BBNv} is close to the helium boundary, may be less than 2. 

In the next section we study the nuclear boundaries without regard to BBN\@.  Since the nuclear boundaries depend on $m_a$ but not $v$ the relevant distribution is
  \be
dP = f_m(m_a) \, n_{nuc} (m_a) \, d \ln m_a,
\label{eq:P4}
\ee
where the effective distribution $f_m(m_a)$ results from integrating (\ref{eq:P2}) over $v$.  When $m_e$ is held fixed, our universe lies very close to the tip of the cone formed by the nuclear boundaries, as shown by the red dot in Figure~\ref{fig:nuclearboundaries}, providing evidence that $y_{u,d}$ do scan and implying that $f_m(m_{u,d})$ increases with $m_u$ so that our universe is typical in the multiverse.  In this case the integral over $m_{u,d}$ in (\ref{eq:P2}) will be dominated by the part of the observer region near the tip of the cone, suggesting that Figure~\ref{fig:BBNv} does indeed provide a good understanding of the environmental selection of $v$.  In Section~\ref{sec:atomic} we detail the analysis that yields Figure~\ref{fig:nuclearboundaries} and we extend the analysis to include scanning of the electron mass.

The helium boundary shown in Figure~\ref{fig:BBNv} corresponds to a slice through parameter space.  While we have argued above that it may be a reasonable approximation after integrating over the nuclear boundaries, a critical question is the shape of the helium boundary surface in the 4d space of $(m_a,v)$.  Hence, in Section~\ref{sec:BBNlowTF} we study BBN with $(m_a,v)$ scanning, so that the relevant distribution is (\ref{eq:P2}).  We restrict our attention to the case that the weak interaction freezeout occurs after the QCD phase transition.   In particular, we compute $Y_4(m_a,v)$ to determine the sensitivity of helium production on $m_a$ in the observer region of the nuclear boundaries.  For the picture of the helium boundary of Figure~\ref{fig:BBNv} to survive, it is important that there are no values of $m_a$ in the nuclear observer window that allow substantial hydrogen to survive as $v$ is made very large.  All other parameters that affect BBN are held fixed in Section~\ref{sec:BBNlowTF}, in particular $M_{\text{pl}}/\Lambda_{\text{QCD}}$ and the baryon asymmetry $\eta$, and we assume that decays of quarks and leptons of the second and third generations do not affect BBN\@.    

The assumption of fixed $M_{\text{pl}}/\Lambda_{\text{QCD}}$ is non-trivial since changes in the masses of the heavy quarks, induced by scanning $v$, affect the running of the strong coupling and hence $M_{\text{pl}}/\Lambda_{\text{QCD}}$, if $\alpha_s(M_{\text{pl}})$ is fixed.  Of course, if $\alpha_s(M_{\text{pl}})$ is fixed the ratio $M_{\text{pl}}/\Lambda_{\text{QCD}}$ also becomes sensitive to the masses of any colored particles in the UV completion of the SM\@.   In Section~\ref{sec:BBNlowTF} we avoid these sensitivities by assuming that other environmental constraints select $M_{\text{pl}}/\Lambda_{\text{QCD}}$ to be close to our observed value.  This is certainly reasonable since many such constraints have been studied~\cite{Graesser:2006ft}.    The effect of heavy quark thresholds on $M_{\text{pl}}/\Lambda_{\text{QCD}}$ is computed in the appendix, together with the size of the corresponding variation in $\alpha_s(M_{\text{pl}})$ that keeps $M_{\text{pl}}/\Lambda_{\text{QCD}}$ fixed.

In multiverse physics, as more parameters are allowed to scan runaway behavior typically results, unless further constraints are imposed.   In Section~\ref{sec:scanetaMP}, we scan $\eta, M_{\text{pl}}$ and find that BBN is no exception.
Reducing $\eta$ or $M_{\text{pl}}$ allows the reaction $np \rightarrow d \gamma$ to freeze-out before all the nucleons are burnt to heavier nuclei, while increasing $M_{\text{pl}}$ changes the condition for neutron-proton freeze out, allowing $v$ to increase as in (\ref{eq:v}).  We briefly discuss physics that could prevent these runaways.

%%%%%%%%%%%%%%%%%%%%%%%%%%%
\section{The Nuclear Boundaries and $m_{u,d,e}$}
\label{sec:atomic}
\begin{figure}[h!]
\begin{center} \includegraphics[width=0.5 \textwidth]{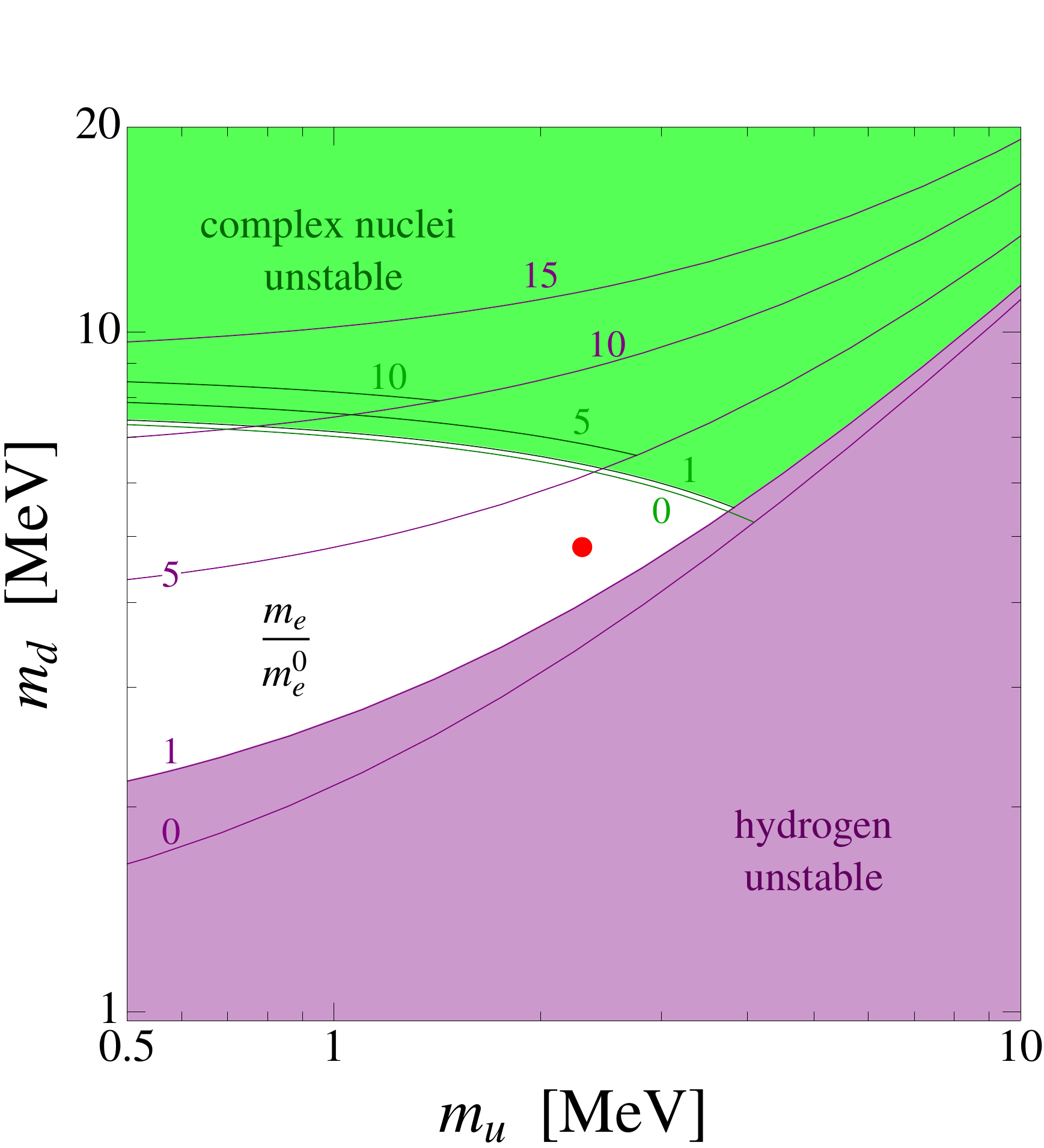}\end{center}
\caption{ \label{fig:ude}
Anthropic requirements in the $(m_u, m_d)$ plane for the stability of complex nuclei (upper, green) and hydrogen (lower, purple) for several values of the electron mass.  The shading corresponds to the excluded regions for $m_e = m_e^0$.  For other values of $m_e / m_e^0$, the allowed region in $(m_u, m_d)$ is the wedge between like-numbered green and purple contours.  For $m_e \gtrsim 15 m_e^0$, there are no values of the up and down quark masses satisfying the anthropic requirements.
}
\end{figure}

\begin{figure}[h!]
\begin{center} \includegraphics[width=1.0 \textwidth]{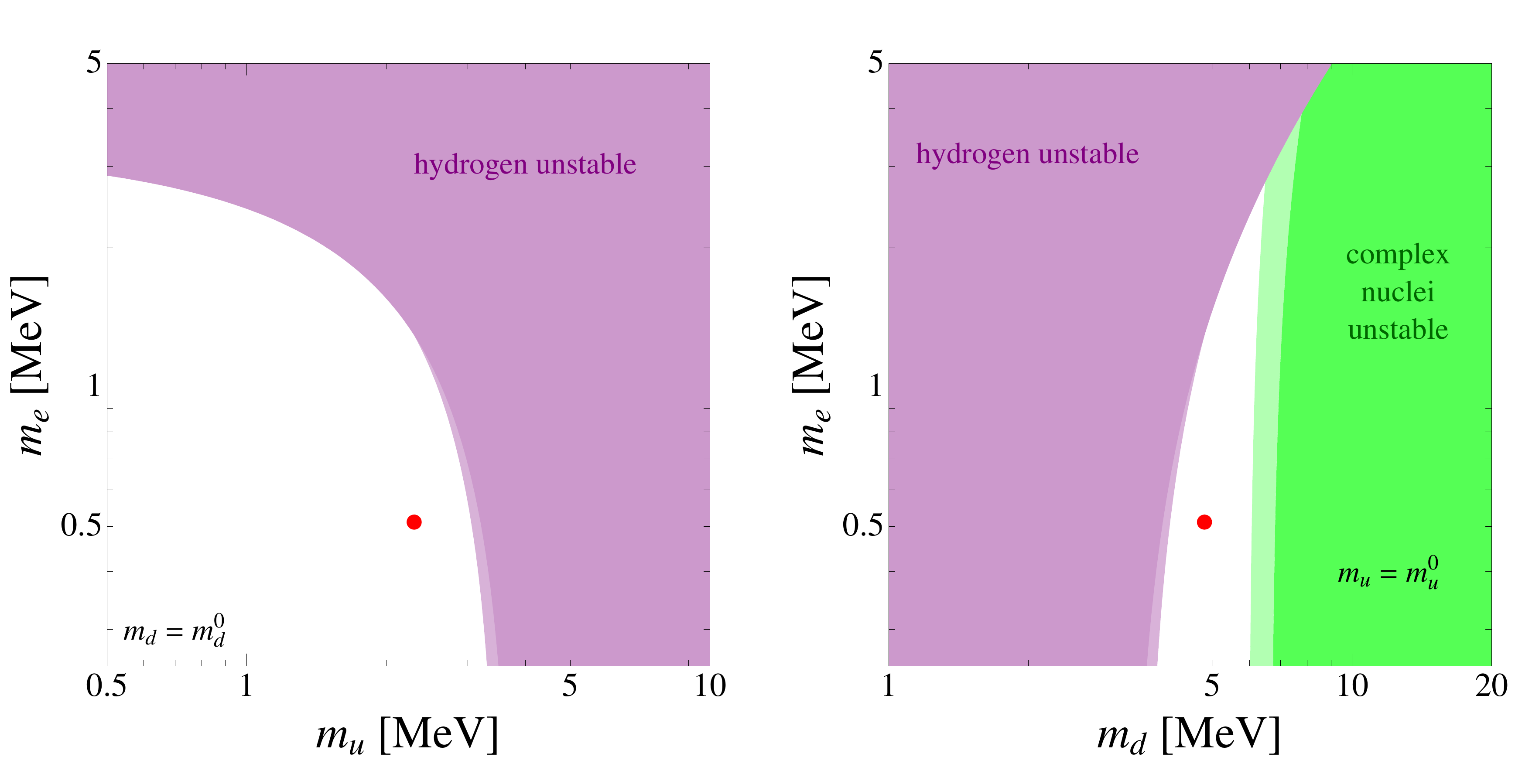}\end{center}
\caption{ \label{fig:ue-de}
Anthropic requirements for the stability of complex nuclei (green) and hydrogen (purple) in the $(m_u, m_e)$ plane at $m_d = m_d^0$ (left) and in the $(m_d, m_e)$ plane at $m_u = m_u^0$ (right).}
\end{figure}

%%%%%%%%%%%%%%%%%%%%%%%%%%%%
 The quarks and charged leptons have masses that vary over about six orders of magnitude.  These masses are hierarchical, and while the quark mixing angles are small, those of the lepton sector are much larger.  The origin of these masses is not clear and could, for example, result from a combination of symmetry and statistics. Here we stress that if the origin is entirely from symmetry, then it is surprising that $(m_u,m_d,m_e)$ lie so close to the boundaries at which complex nuclei and hydrogen are unstable, as shown in Figures~\ref{fig:nuclearboundaries}, \ref{fig:ude}, and~\ref{fig:ue-de}.  Figure~\ref{fig:ude} shows these boundaries in the $(m_u,m_d)$ plane with contours for various values of $m_e$ allowing a visualization of the whole 3d space. Figure~\ref{fig:ue-de} shows the boundaries in other 2d slices: the $(m_u,m_e)$ plane at $m_d = m_d^0$ and the $(m_d,m_e)$ plane at $m_u = m_u^0$.  In all cases it is clear that our universe is strikingly close to the boundaries; $m_d$ is virtually determined by the boundaries, whereas $m_e$ and $m_u$ are near their maximum allowed values, but could have been smaller.
 
It is most convenient to rotate axes and consider our location with respect to these nuclear boundaries in terms of $m_d \pm m_u$.  As the up-down mass splitting is taken small, the neutron-proton mass difference decreases approximately linearly,
\be\label{eq:npmass}
m_n - m_p \approx \delta_{\text{iso}} \frac{\left(m_d - m_u\right)}{\left(m_d - m_u\right)_0} + \delta_{\text{EM}}.
\ee
We take the isospin-violating contribution to the neutron-proton mass difference from the lattice~\cite{Walker-Loud:2014iea}, $\delta_{\text{iso}} = 2.39 \pm 0.21$ MeV; the electromagnetic contribution is then determined by the measured value of $m_n - m_p \approx 1.293$ MeV\@.  Once the neutron-proton mass difference drops below the electron mass, hydrogen becomes unstable to electron capture by the proton.  This is the purple-shaded region in Figures~\ref{fig:nuclearboundaries}, \ref{fig:ude}, and~\ref{fig:ue-de}, with light purple denoting the $1 \sigma$ excluded region.

In the orthogonal direction, the square of the pion mass increases linearly with the sum of the up and down masses, decreasing nuclear binding energies.  If the binding energy per nucleon, $B/A$, is sufficiently small,
\be
|B/A| < m_n - m_p - m_e,
\label{eq:heavnucstab}
\ee
then neutrons bound in the nucleus will decay~\cite{Agrawal:1997gf}.  We take the pion mass dependence of nuclear binding energies from~\cite{Damour:2007uv}, but it should be noted that our boundary, (\ref{eq:heavnucstab}), is parametrically stronger than the one used in that paper (the complete absence of heavy nuclear bound states, $B/A > 0$).   The green-shaded region in Figures~\ref{fig:nuclearboundaries}, \ref{fig:ude}, and~\ref{fig:ue-de} corresponds to the limit from stability of $^{16}$O, but the location of the boundary is roughly independent of atomic number.  We show the $1\sigma$ uncertainty shaded in light green in Figures~\ref{fig:nuclearboundaries}, \ref{fig:ude}, and~\ref{fig:ue-de}, in which the dominant error arises from the extrapolation of nuclear binding energies away from the SM value of the pion mass~\cite{Damour:2007uv}.  

Consider the variation of $m_{u,d}$ arising from holding $y_{u,d}$ fixed and varying $v$.  This direction is shown as a dashed line in the right panel of Figure~\ref{fig:nuclearboundaries}.  Increasing the quark masses by $\left(26 \pm^{9}_6\right) \%$ (at $1\sigma$ uncertainty) results in unstable complex nuclei, as described above.  Our proximity to this boundary was the original motivation for anthropic solutions to the hierarchy problem~\cite{Agrawal:1997gf}.  However, taking the perspective that, {\it a priori}, we might have lived at any point in the allowed region, then all parallel lines should be equally likely.  While the multiverse force toward large $v$ would explain our proximity to the green region, it is surprising that we should be so close to the tip of the cone.  Furthermore, the electron mass is within a factor of $\sim 2.5$ of the tip in the 3d space of $m_{u,d,e}$.   Our universe could have been orders of magnitude away from this tip in both $m_u$ and $m_e$, while still being close to the complex nuclei boundary.  Thus our position in the 3d space of $m_{u,d,e}$ cannot be explained by a multiverse force to large $v$ alone.

In an ensemble of the SM with quark masses varying, as detailed in~\cite{Hall:2007ja}, if the distributions are flat in the plane of the left panel of Figure~\ref{fig:nuclearboundaries}, then the probability of finding quarks as close to both atomic boundaries as observed is about $10^{-3}$.  However, some of the pattern of quark masses might result from symmetries, and in ensembles of popular models of flavor this probability can be higher~\cite{Hall:2007ja}.  Even so, it is remarkable that the observed values of $m_{u,d,e}$ lie so close to the tip of a 3d cone formed by just the two boundaries of hydrogen and complex nuclei stability.  Moderate power law distributions favoring large values of $m_u$ and $m_e$ appear to provide the most successful predictions known for the first generation masses.

%%%%%%%%%%%%%%%%%%%%%%%%%%%
\section{BBN with Freezeout below the QCD Scale}
\label{sec:BBNlowTF}
%%%%%%%%%%%%%%%%%%%%%%%%%%%%
We now consider the primordial abundance of helium in the full four-dimensional space $(m_{u,d,e}, v)$, with $v/v_0 < 10^2$, subject to the nuclear boundaries described in the previous section.  We will see that the qualitative behavior shown in Figure~\ref{fig:BBNv}, an exponential increase in the mass fraction of helium as $v$ is raised, holds throughout the space of allowed particle masses, although the quantitative details will be affected.  Crucially, requiring that hydrogen make up at least $10\%$ of the mass fraction of the universe will bound $v$ to within an $\mathcal{O}\left(10\right)$ factor of its present value, whenever $m_{u,d,e}$ are such that we have stable hydrogen and heavy nuclei.

Almost every neutron that is sufficiently long-lived will become bound up in $^4$He.  Thus the most important factor in determining the ultimate abundance of helium is the initial neutron fraction.  At high temperatures, both neutrons and protons are in equilibrium.  However, the reaction that interconverts protons and neutrons freezes out at a temperature
\be
T_{\text{fo}} \approx \left(0.9 \text{ MeV}\right) \left(\frac{v}{v_0}\right)^{4/3} \left(\frac{M_{\text{pl},0}}{M_{\text{pl}}}\right)^{1/3},
\label{eq:Tfo}
\ee
valid whenever $T_{\text{fo}} \gg m_e, m_n - m_p$.\footnote{For the figures, we compute the freeze-out temperature numerically in order to capture the finite-mass effects.}  Below this temperature, the ratio of neutron to proton abundances is given by
\be
\frac{n}{p} \, \approx \, e^{-(m_n-m_p)/T_{\text{fo}}} \; e^{-\Gamma_n t},
\ee
in which $\Gamma_n$ is the neutron width, and we have neglected neutrino degeneracy effects.  For $m_e \ll m_n - m_p$,
\be
\Gamma_n \approx \frac{1}{880 s} \left(\frac{v}{v_0}\right)^4 \left(\frac{\left(m_n - m_p\right)_0}{m_n - m_p}\right)^5.
\ee
Electromagnetic corrections and finite electron mass effects~\cite{Bedaque:2010hr} are included in the figures.  Due to the large entropy-per-baryon in the early universe, deuterium is efficiently photo-dissociated until the temperature drops such that the fraction of photons with sufficient energy to destroy deuterium is of order the baryon-to-photon ratio.  This marks the end of the deuterium bottleneck, at a temperature $e^{B_d/T_d} \sim \eta$.  At this temperature, roughly all remaining neutrons are processed into $^4$He, leading to a helium mass fraction
\be
Y_4 \approx \frac{2(n/p)}{1 + (n/p)},
\ee
with $(n/p)$ evaluated at the end of the deuterium bottleneck.

\begin{figure}[h!]
\begin{center} \includegraphics[width=1.05 \textwidth]{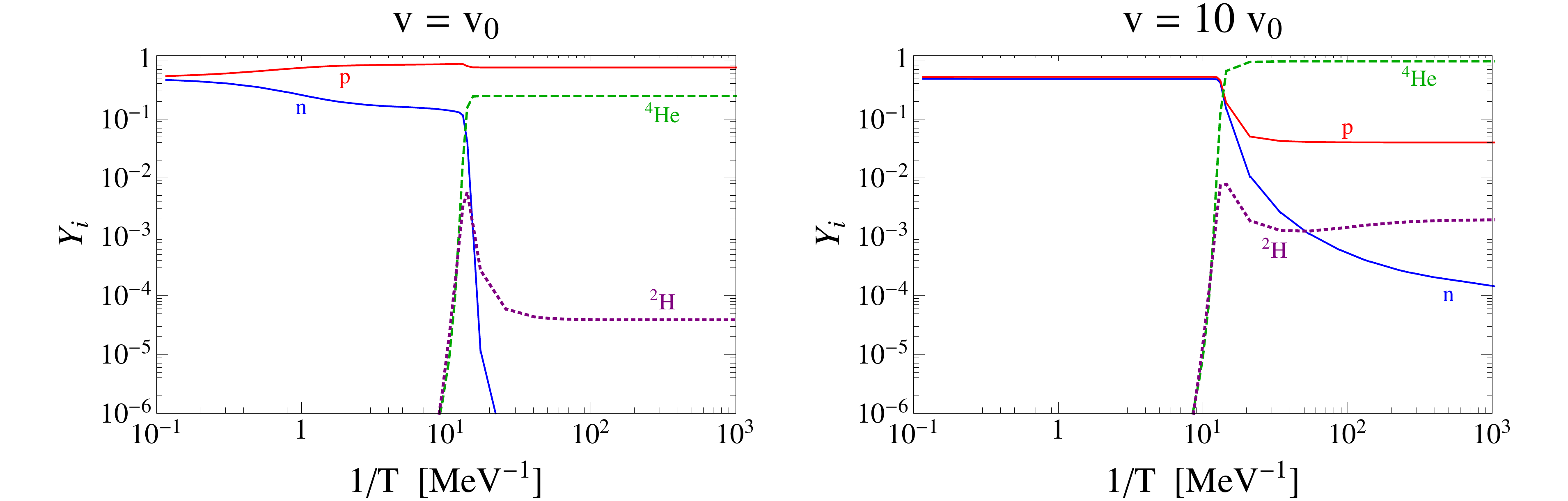}\end{center}
\caption{ \label{fig:history}
The thermal history of proton, neutron, deuterium, and helium mass fractions.  The left panel shows BBN in the Standard Model, while the right shows the effect of raising $v$ to $10 \, v_0$.  All other relevant parameters, such as $m_{u,d,e}$, are fixed to their SM values.
}
\end{figure}  

The left panel of Figure~\ref{fig:history} shows the thermal history of a few relevant elemental abundances in our universe, illustrating the effects described in the preceding paragraph.  As described in the introduction, the initial neutron-to-proton ratio is approximately 1/6 due to an apparent coincidence among $v, M_{\text{pl}},$ and $m_n - m_p$.  Neutron decays before the end of the deuterium bottleneck at $T_d \sim 0.1$ MeV give $(n/p) \approx 1/7$ and thus $Y_4 \approx 0.25$.  The right panel shows the effect of raising $v$ by an order of magnitude, while holding the light quark and lepton masses fixed.  Because $v$ is large, the freeze-out temperature of neutron-proton interconversion is above 10 MeV, and so $(n/p)$ does not evolve with temperature before the end of the deuterium bottleneck.  Thus, the relative abundance of neutrons is much higher on the right than on the left, leading to the increased yield of helium, as shown in Figure~\ref{fig:BBNv} in the introduction.  Furthermore, the neutron lifetime is much longer on the right-hand side of the figure, resulting in a finite abundance of neutrons down to very low temperatures.   It is interesting to note that the abundance of helium flattens off even though there are a few remaining neutrons.  This is due to the freeze-out of the deuterium production reaction, after which point the neutrons can no longer be processed, leading to a finite mass fraction of hydrogen at arbitrarily large $v$.  For SM values of $m_u, m_d,$ and $m_e$, this asymptotic value of $Y_4$ is approximately $98\%$.

The abundances shown in Figure~\ref{fig:history} and throughout the paper were calculated using a modified version of {\tt AlterBBN}~\cite{Arbey:2011nf}, a public BBN code.  The quark mass dependence of the neutron-proton mass splitting is taken from lattice results~\cite{Walker-Loud:2014iea}, as described in Section~\ref{sec:atomic} for the nuclear boundaries, and is included in both the neutron lifetime and the neutron-proton interconversion rate.  The pion mass dependence of the deuterium binding energy and scattering length is taken from~\cite{Berengut:2013nh}, in which it is found that deuterium remains bound throughout the region of interest.  For the deuterium production cross-section, we include only the leading order isovector magnetic exchange, as it is the dominant contribution in the temperature range of interest~\cite{Rupak:1999rk}.  The variation of other nuclear binding energies and reaction rates has only a minor effect on the final helium abundance~\cite{Berengut:2013nh} and is therefore neglected.  Furthermore, we neglect the minor electron mass dependence of the time-temperature relationship and its effect on the neutrino decoupling temperature, having checked that they do not appreciably affect the results.

\begin{figure}[h!]
\begin{center} \includegraphics[width=1.00 \textwidth]{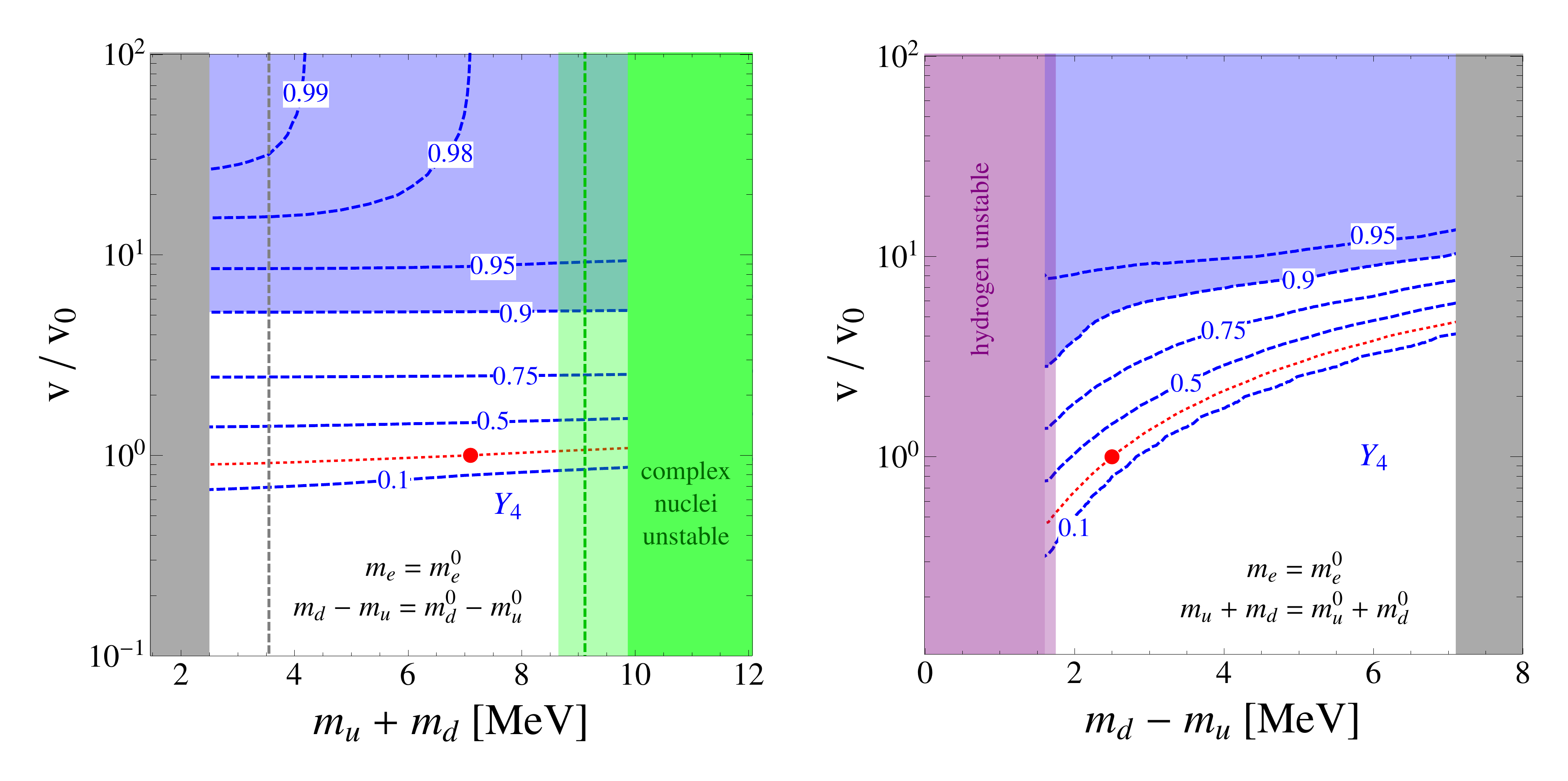}\end{center}
\caption{ \label{fig:Ycontours}
The helium mass fraction as a function of $v$ and the quark masses, fixing the electron mass to its SM value.  The blue shaded region has a helium mass fraction greater than 90\%.  In the left panel $m_d-m_u$ (and therefore the neutron-proton mass splitting) is held fixed, while $m_d+m_u$ (and hence the pion mass) is held fixed on the right.  The nuclear boundaries corresponding to instability of complex nuclei and hydrogen are shown in green and purple in the left and right panels, respectively, with translucent shading corresponding to $1\sigma$ uncertainties.  The calculation of the deuterium binding energy and scattering length in~\cite{Berengut:2013nh} cuts off at the dashed gray line.  Below this point, we linearly extrapolate their results.  We shade in gray the region with $m_u < 0$, since we take $m_u$ to be positive.  The red dot and contour correspond to SM parameter values and helium mass fraction, respectively.
 }
\end{figure}

Figure~\ref{fig:Ycontours} shows how the mass fraction of helium depends on $v$ as a function of $m_d+m_u$ (left) and $m_d-m_u$ (right).        The sum of the quark masses is proportional to the square of the pion mass, while the difference is linearly related to the proton-neutron mass splitting (\ref{eq:npmass}).  Once the pion mass becomes $\sim 10-18\%$ larger than in the SM, complex nuclei become unstable to beta decay, shown shaded in green in the figure.  In principle, decreasing the quark masses may cause the dineutron and diproton to become bound, as the inter-nucleon forces are strengthened.  However, theoretical errors on the dinucleon scattering length are large~\cite{Berengut:2013nh,Beane:2002vs}, so that it is as yet unclear whether this occurs.  Nevertheless, we expect our results to be largely unaffected by this possibility.  While the dineutron could in principle provide a mechanism for sequestering neutrons or converting them into protons via decays to deuterium, it seems likely that there should not be an appreciable dineutron abundance during BBN\@.  Until a temperature $T_{\text{nn}} \sim B_{\text{nn}} \log\eta$, where $B_{\text{nn}}$ is the putative dineutron binding energy, the dineutron abundance should be highly suppressed by photodissociations, in exact analogy with the deuterium bottleneck.  BBN proceeds quickly after the end of the deuterium bottleneck, so that, as long as the dineutron binding energy is not too close to that of deuterium, the dineutron abundance will remain negligible throughout BBN\@.  By the time dineutrons can form without becoming immediately photodissociated, nearly all of the available neutrons will have become bound up in helium.

\begin{figure}[h!]
\begin{center} \includegraphics[width=0.5 \textwidth]{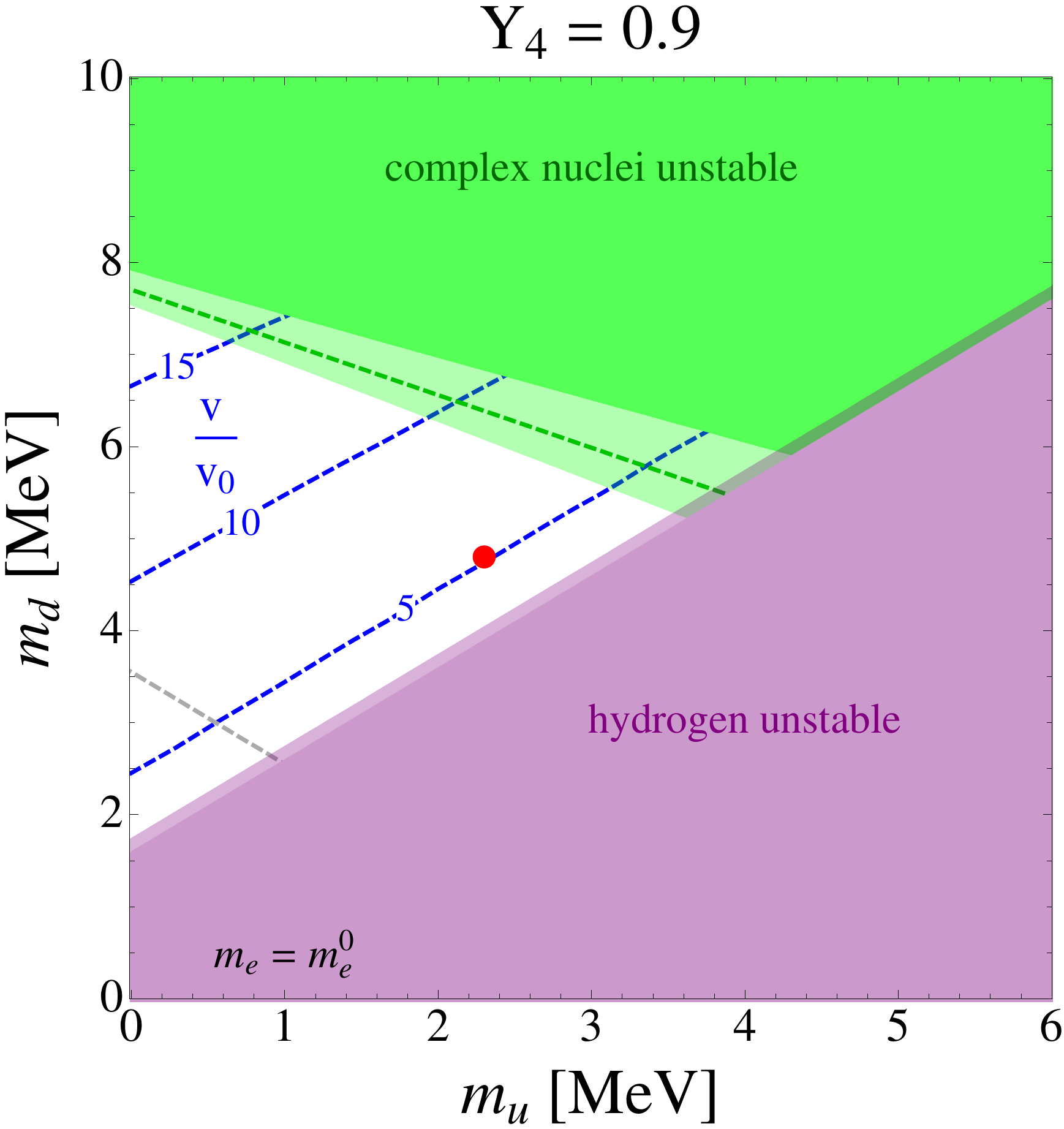}\end{center}
\caption{ \label{fig:BBNvmumd}
Contours of $v$ relative to its SM value in the $(m_u, m_d)$ plane corresponding to $Y_4 = 0.9$.  Here the electron mass is fixed to its value in the Standard Model.  Colored regions (and the dashed gray contour) are identical to those in Figure~\ref{fig:Ycontours}.  The red dot shows the SM values of $(m_u,m_d)$; at this point the universe becomes dominated by helium when $v$ is raised by a factor of approximately 5.  Throughout the region allowed by the nuclear boundaries, $v$ cannot be raised to more than $(15-20) \, v_0$ before the universe becomes dominated by helium.
}
\end{figure}

\begin{figure}[h!]
\begin{center} \includegraphics[width=1.0 \textwidth]{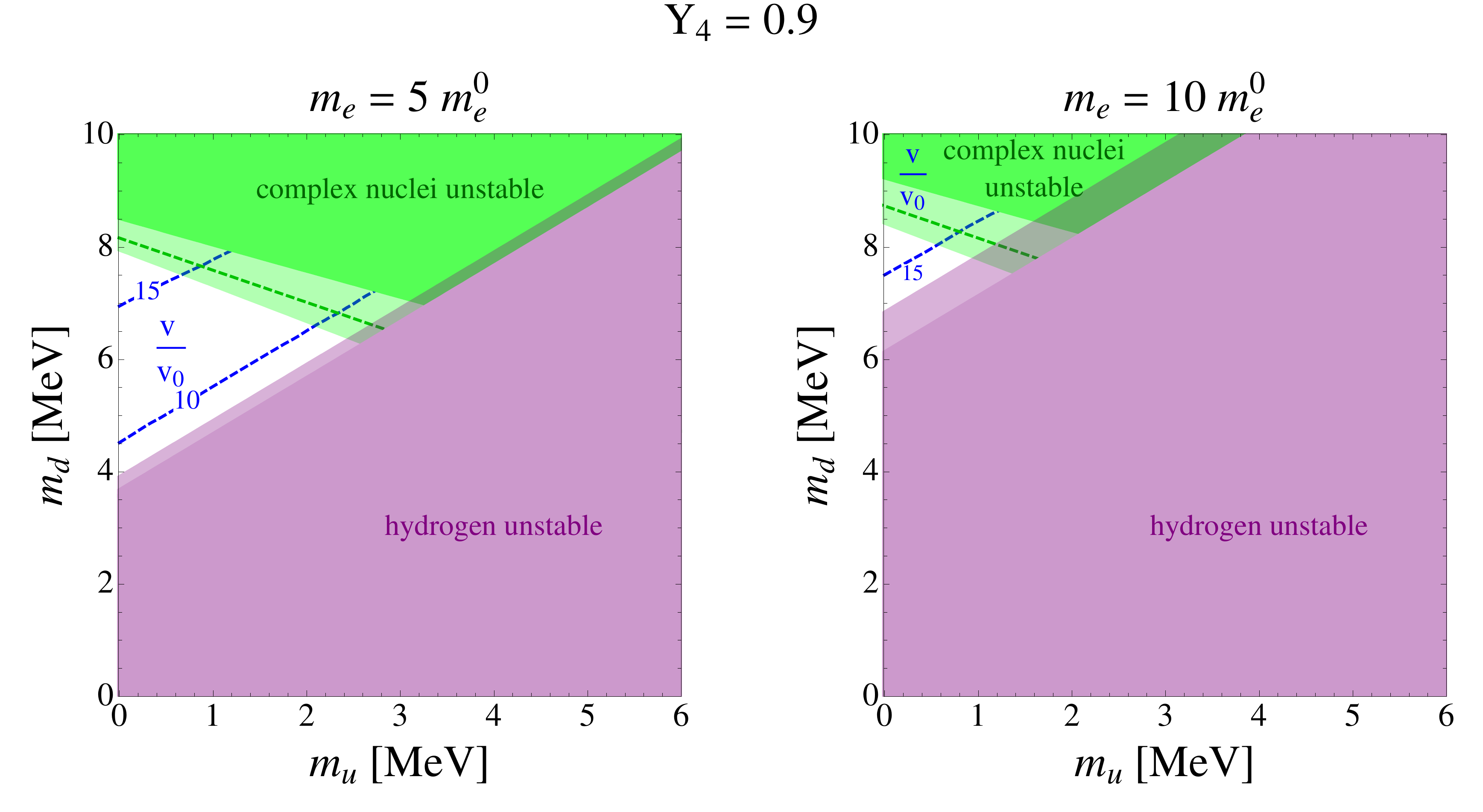}\end{center}
\caption{ \label{fig:BBNvmumdme}
As Figure~\ref{fig:BBNvmumd}, but with $m_e$ fixed to 5 (10) times its SM value in the left (right) panel.  Even at large values of the electron mass, $v$ remains bounded by the requirement that $Y_4$ not exceed $90\%$.
}
\end{figure}

Conversely, diprotons present an additional channel for producing helium, converting protons to neutrons via decays to deuterium and thereby potentially increasing the $^2$H abundance (and thence $Y_4$) relative to BBN in the Standard Model~\cite{MacDonald:2009vk}.  However, other estimates~\cite{Bradford:2009} suggest that photodissociation of diprotons is important until after diproton production freezes out, resulting in a negligible diproton abundance during BBN\@.  In either case, the existence of the diproton bound state is unlikely to qualitatively affect our conclusions, since it will not deplete the mass fraction of helium formed during BBN\@.  The curves shown in Figures~\ref{fig:Ycontours}-\ref{fig:BBNvmumdme} may then be considered a lower bound on the amount of helium production.

Within the allowed region, the helium fraction depends only weakly on the pion mass.  By virtue of the initial $(n/p)$ ratio, $Y_4$ increases rapidly with $v$ until it begins to saturate due to the freeze-out of $n p \rightarrow d \gamma$.  The value of $Y_4$ at this point depends on reaction rate, and therefore on the pion mass.  Increasing the pion mass causes the reaction to freeze out earlier, reducing the asymptotic yield of helium at large $v$.  So long as the anthropic bound on the mass fraction of helium is below about $95\%$, however, our results are independent of this effect.

Fixing the value of the pion mass to its value in the Standard Model, $Y_4$ depends on the difference between the light quark masses through two major effects.  For small values of the mass difference, the dominant effect is once again the initial $(n/p)$ ratio.  Since the freeze-out temperature for neutron-proton interconversion scales like $T_{\text{fo}} \sim v^{4/3}$, contours of constant $Y_4$ correspond to contours of $v \sim (m_n - m_p)^{3/4}$.  At larger values of $m_d - m_u$, however, the phase-space suppression of neutron decay is quickly alleviated, and neutrons become short-lived relative to the end of the deuterium bottleneck.  In order to raise the lifetime, thereby ensuring a large enough supply of neutrons to create $^4$He, $v$ must again be increased along contours of constant $Y_4$.  

Fixing to a particular value of $Y_4$, we can view the entire $(m_u, m_d, v)$ space at fixed electron mass.  Since we do not know the absolute location of the anthropic boundary in $Y_4$, we (arbitrarily) fix the helium mass fraction to $90\%$.  Figure~\ref{fig:BBNvmumd} shows contours of $v$ corresponding to $Y_4 = 0.9$ in the $(m_u, m_d)$ plane, with $m_e$ fixed to its SM value.  Once again, hydrogen instability is shaded in purple and complex nuclei become unstable to neutron decay in the green shaded region at large $m_\pi$.  At the red dot, corresponding to the SM values of $m_u$ and $m_d$, the helium mass fraction reaches 0.9 when $v$ is raised to approximately 5 $v_0$.  The contours of $v$ increase in the direction of increasing $m_d - m_u$, and therefore $m_n-m_p$, due to the neutron lifetime and the initial $(n/p)$ ratio, as described in the previous figure.  This increase is cut off by the region where complex nuclei are unstable, so that $v$ remains bounded in the full $(m_u,m_d)$ plane.

The situation remains quite similar under variation of the electron mass.  Figure~\ref{fig:BBNvmumdme} shows the same contours of $Y_4 = 0.9$ as in Figure~\ref{fig:BBNvmumd}, but with $m_e$ fixed to larger values.  Due to the dependence of the nuclear boundaries on the electron mass, larger values of both the pion mass and $m_d - m_u$ become allowed as the electron mass is increased.  The larger pion mass is responsible for the curvature of the $v$ contours.  As shown in Figure~\ref{fig:Ycontours}, $Y_4$ begins to become independent of $v$ as it reaches its asymptotic value.  This value decreases as the pion mass is increased.  In the orthogonal direction, the effect of the increased neutron-proton mass splitting on the neutron lifetime is mitigated by the larger electron mass, so that significantly larger values of $v$ are not necessary to maintain $Y_4 = 0.9$.  Furthermore, as the electron mass is taken above $m_e \gtrsim v^{4/3} M_{\text{pl}}^{-1/3}$, electrons become non-relativistic at temperatures relevant for the freeze-out of neutron-proton interconversion.  The interconversion rate is proportional to electron number density and so will freeze out at $T_\text{fo} \sim m_e$.  Thus, as the electron mass is raised, the helium mass fraction is increased at fixed $v$.  The end result is that requiring the existence of stable hydrogen, complex nuclei, and a helium mass fraction below $90\%$ bounds the electroweak VEV to within a factor of about 20 above its Standard Model value.
  
%%%%%%%%%%%%%%%%%%%%%%%%%%%
\section{Scanning $\eta$ and $M_{\text{pl}}$}  
\label{sec:scanetaMP}
%%%%%%%%%%%%%%%%%%%%%%%%%%%

In the previous section, we showed that our universe lives within a finite volume in the $(v, y_u, y_d, y_e)$ space, bounded by the helium wall and the nuclear physics boundaries.  As a next step, it is interesting to ask if this result is robust to varying even more parameters.  The BBN helium abundance depends on both the baryon to photon number density, $\eta$, and the Planck scale, $M_{\text{pl}}$ and in this section we study whether varying these parameters allows $v$ to runaway to large values while preserving a significant hydrogen fraction.  For $v/v_0$ larger than about 100, freezeout occurs before the QCD phase transition leading to a new regime for the calculation of $Y_4$, as shown in Figure~\ref{fig:BBNv} and discussed in detail in Section~\ref{sec:BBNhighTF}.  In this section we consider the possibility of runaway of $v$ to values of order $100 \, v_0$.   We assume $m_{u,d,e}$ are selected by the nuclear physics boundaries as in Section~\ref{sec:atomic} to values close to those we observe and hence, for simplicity, we take them fixed at the observed values.

It is not surprising that varying more parameters leads to runaway directions, because generically a new dangerous wall is required each time the dimensionality of the parameter space is increased by 1, in order to contain a finite volume.  The existence of runaway directions indicates that new walls must be identified, and/or the prior distributions of the parameters must disfavor the runaway directions (trivially, some variables may not scan in the multiverse).  Below, we give preliminary remarks about how extra walls or prior distributions may stop the runaways we identify.

\begin{figure}[h!]
\begin{center} \includegraphics[width=1.0 \textwidth]{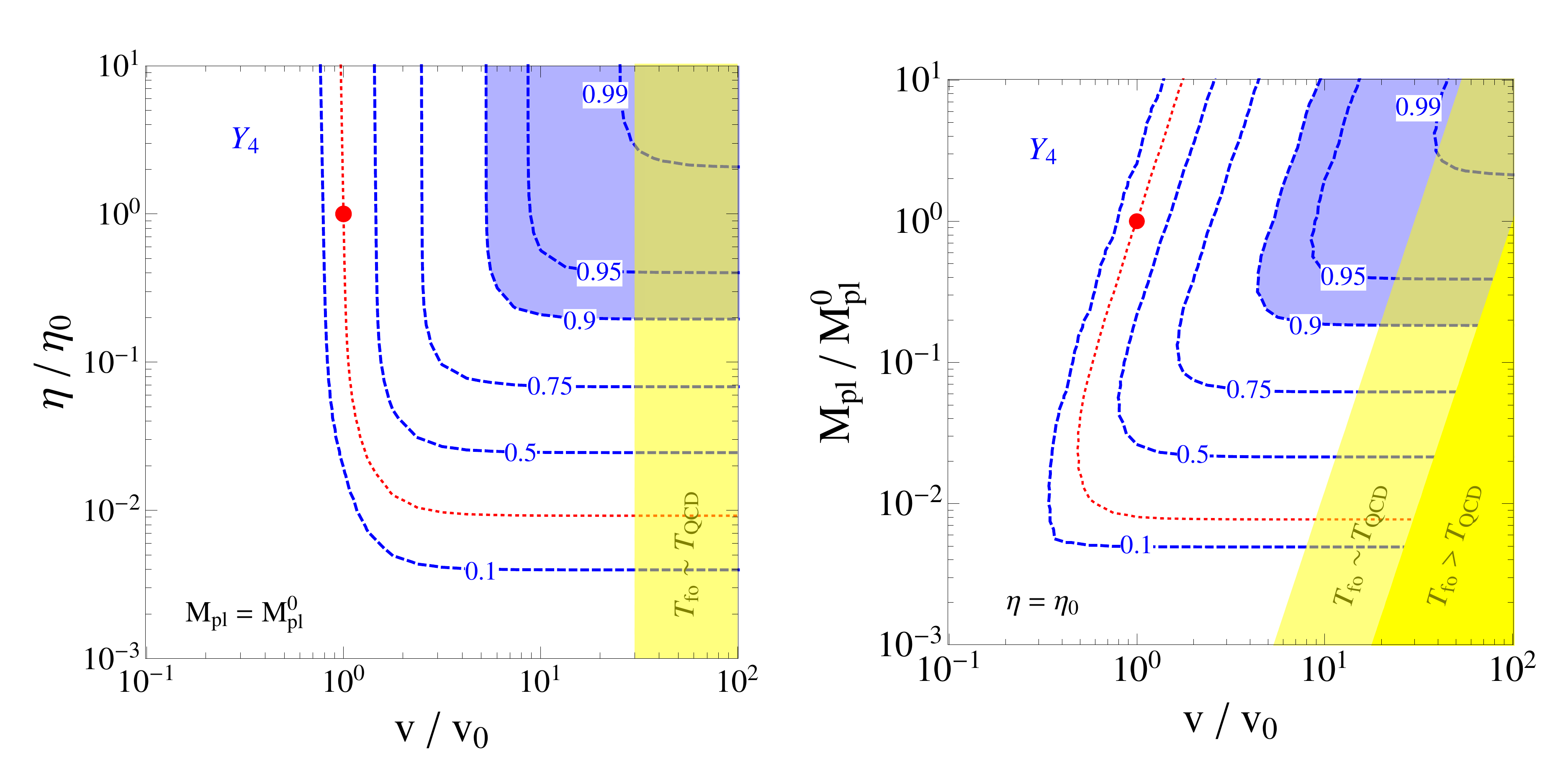}\end{center}
\caption{ \label{fig:cosrun}
The helium mass fraction, $Y_4$, is shown in the $(v, \eta)$ plane to the left and the $(v, M_{\text{pl}})$ plane to the right, fixing other parameters relevant for BBN, such as $m_{u,d,e}$, to their SM values.  We see that there are possible runaways to large $v$, keeping $Y_4$ small, when $\eta$ is decreased below the observed value, or when $M_{\text{pl}}$ is increased or decreased.  For $v/v_0$ larger than about 100, freezeout occurs before the QCD phase transition leading to a new regime for the calculation of $Y_4$, as discussed in Section~\ref{sec:BBNhighTF}.
}
\end{figure}

We now allow $\eta$ and $M_{\text{pl}}$ to vary, and we determine the impact that these parameters have on BBN\@.   We begin by considering varying the baryon to photon ratio, $\eta$.  The left of Figure~\ref{fig:cosrun} shows contours of the helium mass fraction, $Y_4$, as a function of $v$ and $\eta$.  We see that lowering $\eta$ reduces $Y_4$, and if $\eta$ is lowered by more than two orders of magnitude below the observed value, then $v$ can be significantly increased while maintaining $Y_4 \le Y_4^0$.  

The reason that $Y_4$ is reduced and becomes insensitive to $v$ at lower $\eta$ is that lowering the baryon density causes deuterium production, $p + n \rightarrow d + \gamma$ to become slower than the expansion rate of the universe.  In order to process neutrons into helium, deuterium production must be faster than the expansion rate below the temperature that the deuterium blockade ends, $T_d.$\footnote{Note that the value of $T_d$ is only logarithmically sensitive to the value of $\eta$.}  Otherwise, even if $v \rightarrow \infty$ and the initial $n$ and $p$ number densities are equal, the neutrons and protons never find each other to get processed into helium before the neutrons eventually decay to protons.  Helium is not produced when:
\be \label{eq:NoDeut}
n_b \left< \sigma_d \, v \right> \lesssim H |_{T=T_d},
\ee
which occurs for $\eta$ less than a critical value,  $\eta_c$, given by:
\be \label{eq:eta_c}
\eta_c = \frac{H}{n_\gamma \left< \sigma_d v \right>}\approx 12.5 \frac{1}{M_{\text{pl}} \, T_d  \left< \sigma_d v \right>} \approx 2 \times 10^{-12} \approx 3 \times 10^{-3} \, \eta_0,
\ee
where we used that $\left< \sigma_d \, v \right> \approx 6 \times 10^{-9}~\mathrm{MeV}^2$~\cite{Rupak:1999rk}.  The runaway behavior on the left of Figure~\ref{fig:cosrun} follows from the fact that the observed value of $\eta$ happens to be within a few orders of magnitude of this critical value.  The proximity of $\eta_0$ and $\eta_c$ seems to be a coincidence.\footnote{It would be possible to simultaneously explain Eq.~\ref{eq:v} and why $\eta_0 \gtrsim \eta_c$ if {\it too much} hydrogen is a dangerous wall, however this seems difficult to justify.  A possible reason that primordial helium may be important is the fact that collisional excitation of ionized helium  dominates the cooling of galaxies close to the size of our galaxy~\cite{Katz:1995up}. However helium cooling dominates for a narrow range of galactic masses and even there, removing helium only slows the cooling rate by an order one factor.}

Now we consider the impact on BBN of varying $M_{\text{pl}}$ (see Ref.~\cite{Graesser:2006ft} for an earlier analysis and other possible effects of varying $M_{\text{pl}}$).  We often think of $M_{\text{pl}}$ as being fixed, but among the three dimensional parameters $v, M_{\text{pl}}, \Lambda_{\text{QCD}}$ we can always choose one parameter to be fixed and use it to set the units of the other two parameters.  It is helpful to think of our BBN analysis as occurring at fixed $\Lambda_{\text{QCD}}$ with $v$ and $M_{\text{pl}}$ defined in units of $\Lambda_{\text{QCD}}$, in which case varying $M_{\text{pl}}$ corresponds to varying the ratio $M_{\text{pl}} / \Lambda_{\text{QCD}}$.  See appendix~\ref{app:LQCD} for other parameterizations and for the effect of heavy quark mass thresholds on the ratio $M_{\text{pl}} / \Lambda_{\text{QCD}}$.  

The impact of varying $M_{\text{pl}}$ on the helium mass fraction is shown to the right of Figure~\ref{fig:cosrun}.  We can see two possible runaway directions to large $v$.  First, if $M_{\text{pl}}$ is lowered by two or more orders of magnitude, then $v$ can be significantly raised keeping the helium abundance lower than the observed value.  Second, $v$ can be taken large by raising $M_{\text{pl}}$, although to keep $Y_4$ fixed requires $M_{\text{pl}}$ to increase very rapidly as $v^4$.  The runaway at low values of $M_{\text{pl}}$ follows from deuterium production going out of equilibrium, similarly to the above discussion on lowering $\eta$.  As $M_{\text{pl}}$ is lowered, the Hubble expansion speeds up, causing deuterium production to go out of equilibrium ($M_{\text{pl}}$ enters through the relation $H\sim T^2 / M_{\text{pl}}$ in the right side of Eq.~\ref{eq:NoDeut}).  The runaway at large values of $M_{\text{pl}}$ corresponds to a lowering of the temperature where $n \leftrightarrow p$ interconversions go out of equilibrium, as can be seen by the $M_{\text{pl}}$ dependence of Eq.~\ref{eq:Tfo}, depleting the neutron number density and preserving hydrogen.

\begin{figure}[h!]
\begin{center} \includegraphics[width=1.0 \textwidth]{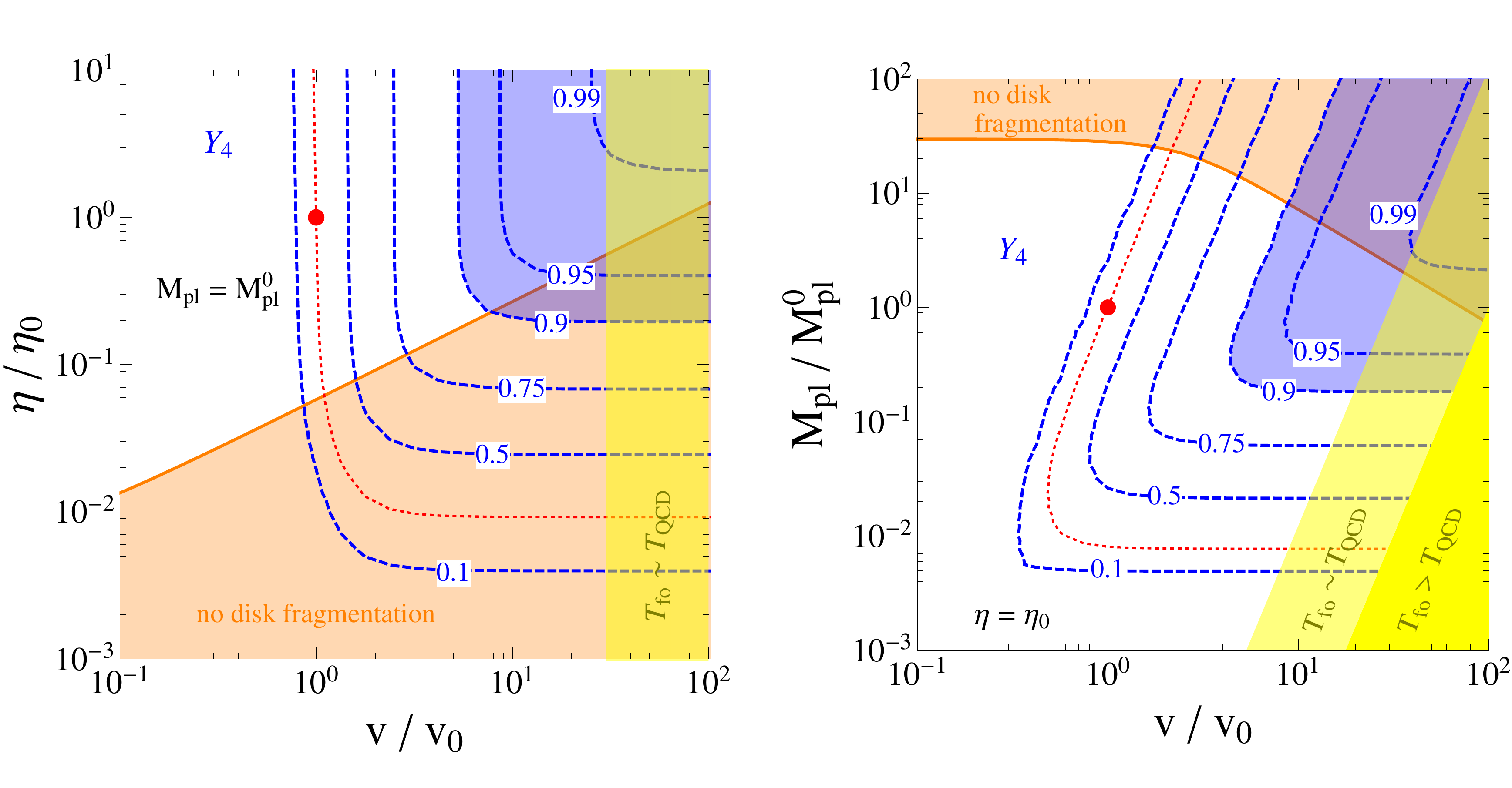}\end{center}
\caption{ \label{fig:cosrunDF}
The helium mass fraction, $Y_4$, versus v, $\eta$, and $M_{\text{pl}}$, as in Figure~\ref{fig:cosrun}, except also including the boundary from disk fragmentation.  Dark matter is assumed to be a WIMP with energy density that scales as $\Omega_{dm} \sim v^2$.  In the orange region, baryonic disks fail to fragment into stars for galaxies the size of our galaxy $M = 10^{12} \, m_\odot$.  We see that the disk fragmentation boundary blocks the runaways to large $v$ at small $\eta$ and at large $M_{\text{pl}}$.  Note that when varying $M_{\text{pl}}$ we are using units where $\Lambda_{\text{QCD}}$ is fixed; the physically relevant quantity is $M_{\text{pl}} / \Lambda_{\text{QCD}}$.
}
\end{figure}

The above analysis shows that once $\eta$ and/or $M_{\text{pl}}$ are varied from their observed values, BBN alone no longer restricts the weak scale to a value near what we observe.  One possibility that prevents this runaway behavior is that $\eta$ and $M_{\text{pl}}$ either do not scan or have prior distributions sufficiently peaked to prevent runaways.  Another possibility, that we now consider, is that there are other dangerous walls, beyond the BBN and nuclear boundaries discussed so far in this paper, that depend on $\eta$ and/or $M_{\text{pl}}$ and block the runaway behavior.

Suppose, for example, that dark matter is a WIMP with annihilation rate that is related to the weak scale, such that the energy density of dark matter increases with the weak scale, 
\be \label{eq:WIMP}
 \sigma_{d}  = \frac{\alpha_d}{4 \pi v^2} \qquad \qquad \rho_d =  \left(\frac{v}{v_0} \right)^2 \left( \frac{M_{\text{pl}}}{M_{\text{pl}}^0} \right)^{-1} \rho_d^0,
\ee
where $\alpha_d$ is a coupling constant defined through the above relation, that is assumed not to scan, and $\rho_d$ is the observed energy density of dark matter, which depends on an inverse power of $M_{\text{pl}}$ because the yield at freeze out is proportional to the Hubble expansion rate.   If the DM energy density becomes too large, relative to the baryon energy density, then baryonic discs within galaxies may not fragment to form stars.  Ref.~\cite{Tegmark:2005dy} finds this constraint to be,
\be \label{eq:DF}
\frac{\rho_b}{\rho_b^0} \, >  \, 0.014 \, \frac{M_{\text{pl}}}{M_{\text{pl}}^0} \, \left(  \frac{\rho_b + \rho_d }{\rho_b^0+\rho_d^0} \right)^{1/3}.
\ee
Eq.~\ref{eq:DF} depends on $M_{\text{pl}}$ explicitly, on $v$ and $M_{\text{pl}}$ through $\rho_d$ as in Eq.~\ref{eq:WIMP}, and on $\eta$ through the relation $\rho_b = \rho_b^0 \, \eta / \eta_0$.  

We show the location of this disk fragmentation boundary on Figure~\ref{fig:cosrunDF}, assuming WIMP dark matter, as in the above discussion.  We see that this boundary blocks the runaway to large $v$ at small $\eta$ and at large $M_{\text{pl}}$.  There remains a runaway to large values of $v$ at small $M_{\text{pl}}$.  This leads to a new regime of high $T_{\text{fo}}$ discussed in the next section.   Note that the location of the disk fragmentation boundary depends on the size of the galaxy under consideration and on the size of the initial density perturbations~\cite{Tegmark:2005dy, Bousso:2009ks}.  Eq.~\ref{eq:DF} and Figure~\ref{fig:cosrunDF} fix the galactic mass to the size of our Galaxy, $M = 10^{12} \, M_\odot$, and the size of the initial density perturbations to the observed size in our universe.

Disk fragmentation is just one possible physical effect that could prevent runaways with $\eta$ and $M_{pl}$.  There are many other possibilities; for example, the lifetime of main sequence stars is very sensitive to $M_{pl}$, and increasing the energy density of WIMP dark matter while decreasing $\eta$ leads to a dilution of observers~\cite{Bousso:2013rda}.  For increases of $v$ by of order $10^2$ or more, the relevant weak interaction freezes out before the QCD phase transition, so a new regime for computing $Y_4$ is entered and we discuss this regime in the next section.

%%%%%%%%%%%%%%%%%%%%%%%%%%%%%%%%%%%%%%
\section{BBN with Freezeout above the QCD Scale}
\label{sec:BBNhighTF}
%%%%%%%%%%%%%%%%%%%%%%%%%%%%%%%%%%%%%%

In our universe weak interaction freezeout occurs during the hadronic phase, at a temperature near 1 MeV, so the relevant reaction for determining the $n/p$ ratio and hence the helium abundance is $n \nu \leftrightarrow p e$.  However, at very large values of $v$ freezeout occurs in the quark-gluon phase, so that the relevant interactions include $d \nu  \leftrightarrow  ue$ and in this case we find
\be
T_{\text{fo}} \simeq 2 \, \mbox{GeV}  \left(\frac{v/v_0}{300}\right)^{4/3}.
\label{eq:tfoqg}
\ee  

The QCD phase transition is not first order, rather it occurs via a smooth ``analytic crossover", with a critical temperature of around $T_{QCD} \sim 150$ MeV~\cite{Bhattacharya:2014ara}.  Thus we expect the usual analysis in the hadronic phase to be accurate for $v/v_0  < 30$, and the analysis given below in the quark-gluon phase to be accurate for $v/v_0 >100$.  For $30 < v/v_0 < 100$ freezeout occurs as the phase transition proceeds, and we are unable to make a reliable computation.  These regions are illustrated in Figure~\ref{fig:BBNv}.

For freezeout in the quark-gluon phase, the $n/p$ ratio depends on the $u/d$ ratio at freezeout, $x$, and is independent of $m_n - m_p$, or even $m_d - m_u$.  Rather $n/p$ is determined by the baryon and lepton number asymmetries via $x = x(\eta_B, \eta_{e,\mu,\tau})$, which can be computed from the conditions of chemical equilibrium for any spectrum of quarks and leptons. The baryon content of the universe falls into three classes according to the value of $x$:
\be
\frac{n}{p} = \begin{cases} 0, & x>2, \; x<-1:  \;\;\; (p, \pi^+)  \hspace{0.5in} \text{``hydrogen"} \\  
\frac{2-x}{2x-1}, & 1/2 < x < 2: \;\;\;\;\;\;\; (n,p)  \hspace{0.5in} \;\; \text{``H/He or n/He"}\\ 
\infty, & -1<x<1/2: \;\;\; \;  (n, \pi^-)  \hspace{0.5in} \text{``neutron"} \end{cases}  \label{eq:noverp}
\ee
While we require $\eta_B$ to be positive, $x$ may have either sign.  When $x$ is negative there are more antiquarks than quarks in either the up or the down sector.

Universes with $n/p=0$ have no BBN and are ``hydrogen universes".  The absence of primordial helium causes some changes from our own universe, for example in halo cooling, but hydrogen universes are similar to our own and and are expected to have similar numbers of observers.   The universes with $1 < x < 2$ are similar to our own, in the sense that BBN leads to most baryons being in hydrogen/helium.  On the other hand, universes with $1/2 < x < 1$ have $n/p >1$, so that after BBN the baryons are mainly in neutrons/helium.  The ``neutron universes" result from $-1<x<1/2$.   There is a narrow region around $x=1$ where the $n/p$ ratio is sufficiently close to unity that helium dominated universes result, and we have argued that the number of observers in these universes is greatly suppressed. 

Do the neutron universes and the neutron/helium universes contain observers?   The neutron lifetime is of order 
\be
\tau_n \sim 10^{11} \, \text{sec} \left(\frac{v/v_0}{10^2} \right)^4
\label{eq:taun}
\ee
so that if weak interactions freeze out before the QCD phase transitions neutrons are very long-lived, decaying cosmologically at or after the eV era.   Values of $v/v_0 > 10^4$ are likely catastrophic as they give $\tau_n$ greater than the observed age of the universe; neutron and neutron/helium universes will not contain any hydrogen at age $10^{10}$ years, and we assume there is an anthropic cost from the cosmological constant for waiting longer for the neutrons to decay.  Furthermore, universes with $10^3 < v/v_0 < 10^4$ have neutrons decaying after the epoch of population III star formation in our universe, and hence are expected to have less heavy elements compared to our universe.  We do not attempt to estimate the anthropic cost of this.  Finally, universes with $10^2 < v/v_0 < 10^3$, so that $T_{\text{fo}}$ is not far above $\Lambda_{QCD}$, appear likely to contain observers.  Neutrons decay before any star formation; the resulting electrons are rapidly thermalized by inverse compton scattering from the photon background, then the protons cool from interactions with the electrons and hydrogen forms, all at rates faster than the expansion rate at the era of neutron decay.

The $u/d$ ratio at freezeout, $x$, depends on the spectrum of quarks and leptons.  A particularly simple and plausible case has the masses of all quarks and leptons, except for $u,d,e, \nu_i$, larger than $T_{\text{fo}}$ so that only these species are in the thermal bath at the freezeout era. The $\mu$ and $\tau$ lepton asymmetries are carried by $\nu_\mu$ and $\nu_\tau$ and play no role in determining $x$.  The electron asymmetry $\eta_e$ is carried by $(e, \nu_e)$ and the baryon asymmetry by $(u,d)$. Applying the conditions of chemical equilibria and electric charge neutrality yields
\be
x \, = \, \frac{4 \eta_B + 2 \eta_e}{7 \eta_B - 2 \eta_e}.
\label{eq:x1}
\ee
For example, the neutron universe of eq. (\ref{eq:noverp}) results for a very large region of parameter space
\be
\eta_e/\eta_B < - 1/6: \;\;\; \;  (n, \pi^-).    \label{eq:nsimple}
\ee
Since we take $\eta_B$ positive, this is almost the entire region with $\eta_e$ negative.

In the case that the cosmological asymmetries are produced well above the weak scale, electroweak sphaleron processes set the $B+L$ asymmetry to zero.  Nevertheless, $\eta_e$ and $\eta_B$ remain as independent asymmetries.  However, after the early generation of the asymmetries, if there is some era when both Lepton Flavor Violation (LFV) and electroweak sphaleron processes are in thermal equilibrium, then there is a single independent asymmetry associated with the conserved $B-L$ symmetry.  In this case $\eta_e/\eta_B$ takes the special value
\be
\left( \frac{\eta_e}{\eta_B} \right)_{LFV} = -\frac{17}{28} 
\ee
which is less than $-1/6$, so the neutron universe necessarily results. The same result follows from generation of a pure baryon asymmetry followed by electroweak sphalerons removing the $B+L$ component leaving a pure $B-L$ component.

\begin{figure}[h!]
\begin{center} \includegraphics[width=.5 \textwidth]{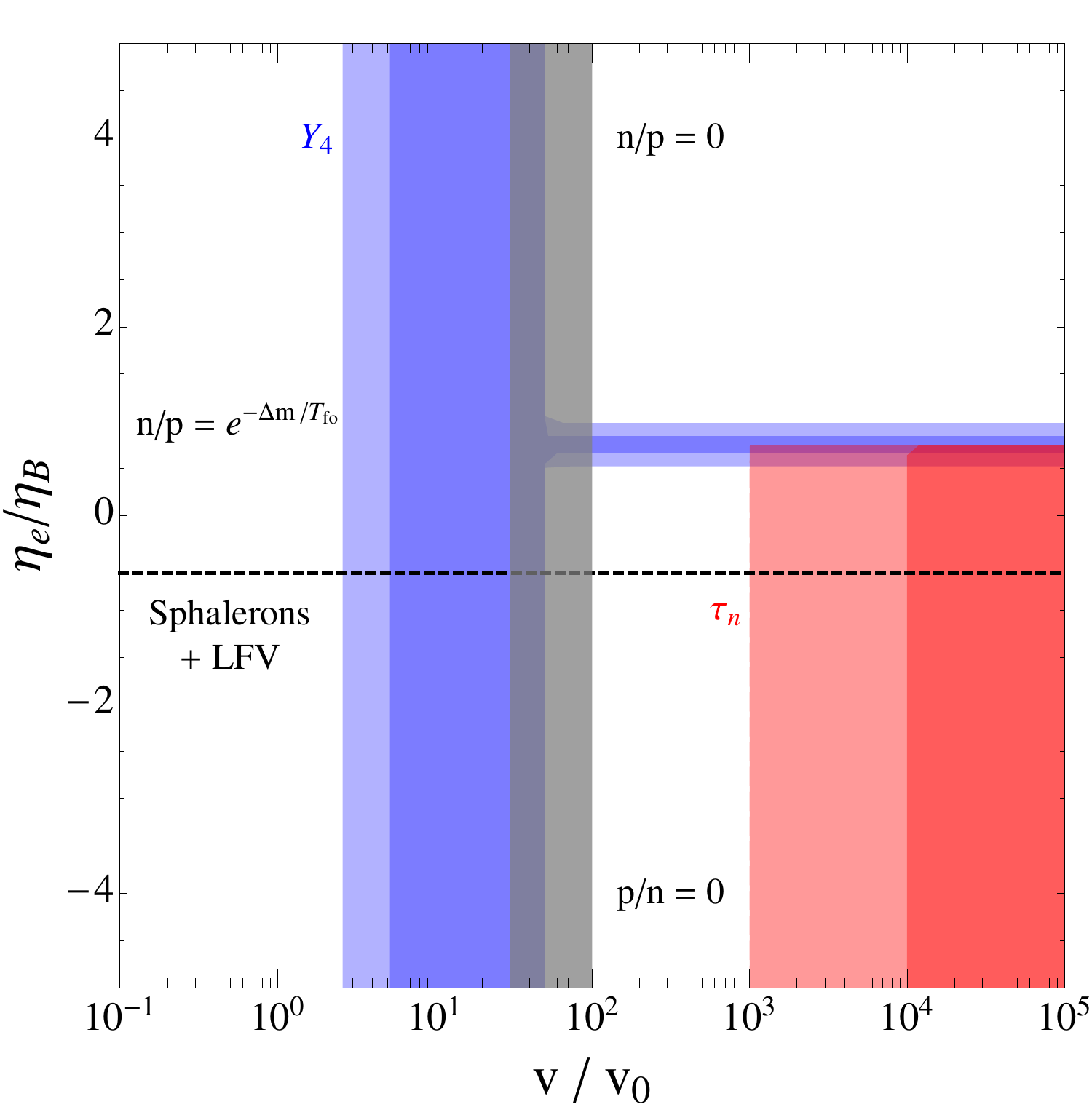}\end{center}
\caption{ \label{fig:phased} Constraints from the helium abundance and the neutron lifetime in the $(v/v_0,\eta_e/\eta_B)$ plane.  The blue shading shows regions with more than $(75, 90)$\% of baryons in helium. For $v/v_0<30$ the helium abundance depends on $v$ but not $\eta_e/\eta_B$, because weak interaction freezeout occurs in the hadronic phase.  The gray shaded region has $30 < v/v_0 < 10^2$; the QCD phase transition is occurring during the era of weak interaction freezeout, so calculations of $n/p$ and $Y_4$ are unreliable. For $v/v_0>100$ weak interaction freezeout occurs in the hadronic phase and the helium abundance becomes independent of $v$ and is large only in a narrow window of $\eta_e/\eta_B$.  The neutron lifetime increases rapidly at large $v$ and leads to a suppression of observers in the red shaded regions, as described in the text.  If there is an era when electroweak sphalerons and lepton flavor violation are both in thermal equilibrium, or if a pure baryon asymmetry is processed by sphalerons, then $\eta_e/\eta_B = -17/28$, as shown by the dashed horizontal line.}
\end{figure}

Bounds on these theories, where heavy flavors of quarks and leptons decouple and are irrelevant, are shown in Figure~\ref{fig:phased}.  The region with helium domination is shown in blue; at $v/v_0 < 30$ this is independent of $\eta_e/\eta_B$, while at $v/v_0 > 10^2$ it is independent of $v$ and helium dominates only in a very narrow band of $\eta_e/\eta_B$ near 3/4.  In the grey shaded region, with $30 < v/v_0 < 10^2$, the QCD phase transition is occurring during the era of weak interaction freezeout, so calculations of $n/p$ are unreliable.  The unshaded regions are anthropically allowed, although in the upper right region there are large numbers of $\pi^+$ that decay very late at large $v$. In the lower right regions, with $v/v_0 > 10^2$ and $\eta_e/\eta_B < 3/4$, the universe contains long-lived neutrons with protons only arising from late neutron decays.  In the dark red region the neutron lifetime is larger than the age of our universe, so that we expect observers to be rare.  In the light red region neutrons decay after the era of population III star formation in our universe, so heavy elements are suppressed.  In the unshaded region with $10^2 < v/v_0 < 10^3$ and $\eta_e/\eta_B < 3/4$ we expect observers, even though protons first appear at or after the eV era.

At sufficiently large values of $v$ the cosmological helium abundance is independent of $v$ and, except for a narrow range of $x$, there is no anthropic helium constraint on $v$.   This does not invalidate the main claim of this paper: the observed value of the weak scale has an anthropic explanation resulting from the rapid dominance of helium as $v$ increases above $v_0$, as illustrated by the red dot of Figure~\ref{fig:BBNv} lying on the steep part of the helium abundance curve.  The parameters of our universe lie very close to this helium boundary, but are quite distant from regions of the landscape where the helium abundance is determined in the quark-gluon phase by $(\eta_B, \eta_{e,\mu,\tau})$.  

Nevertheless, our understanding of $v$ from the nearby helium boundary does require that observers such as ourselves are more probable in the multiverse than observers in universes with freezeout occurring during the quark-gluon phase, even though such universes have much larger values of $v$.    This requirement can be satisfied in two ways:
\begin{enumerate}
\item  To explain our proximity to the nearby helium boundary the effective probability distribution of eq. (\ref{eq:P3}) must have a positive gradient at $v \sim v_0$, so that the probability force in this region is positive $F_v >0$.  However, if $F_v <0$ for $v>10^2 \, v_0$, universes having helium determined by cosmological asymmetries may be less probable than our own.  This change of sign in $F$ may arise from selection effects rather than from the a priori distribution. In regions of $v$ where the weak scale is fine-tuned, $f_v(v)$ contains a factor $v^2$ from the fine-tuning required for electroweak symmetry breaking.  If new physics, such as supersymmetry, cuts off the fine-tuning at scale $m$ between $v_0$ and $10^2 \, v_0$, then $F_v$ decreases by 2 at $m$, and this may be sufficient to change the sign of $F_v$.  This mechanism implies that the new physics which cuts off the fine-tuning is not far above $v_0$ -- a Little Hierarchy is required, not a Large Hierarchy. 
\item The weak scale may affect the environment for observers by changing the physics of stars, for example via the $pp$ reaction in main sequence stars, and by changing the strength of shock waves in supernova explosions that eject heavy elements.  It may be that these have a mild effect on observer selection in the region $v \sim v_0$, where the selection effects from helium dominate, but greatly suppress observers at large values of $v$ where the helium abundance is independent of the weak scale. 
\end{enumerate}

The first possibility predicts new physics well below 10 TeV that cuts off fine-tuning in $v$ and is very exciting for the LHC and future high energy colliders.  The second requires exploration of varying the Fermi coupling in stellar and supernova physics, in particular one would need to show that decreasing the Fermi coupling by four or more orders of magnitude is catastrophic.  

There may be other ways to make the large $v$ universes less probable, but these seem less likely to us. The baryon and lepton asymmetries of the universe may themselves scan, so that for $v> 10^2 \, v_0$ the $n/p$ ratio scans as in eqs. (\ref{eq:noverp}).  The resulting distributions for $n/p$ may make observers with $v> 10^2 \, v_0$ far less common than those with $v \sim v_0$, even if $F_v$ is positive.  For example, it may be that the probability distributions for $(\eta_B, \eta_{L_i})$ strongly favor $-1< x < 1/2$ so that, for $v > 10^2 \, v_0$, BBN results in pure neutron universes.  For $v> 10^4 \, v_0$, neutrons are cosmologically stable, so no observers form.  For $10^2 < v/v_0 < 10^4$ one must study the formation of structure with very long-lived neutrons to assess the likelihood of observers.  Although such universes are clearly very different from ours, for $10^2 < v/v_0 < 10^3$ neutrons decay sufficiently early, and the resulting electrons and protons cool sufficiently rapidly, that large scale structure and star formation appear to form in a very similar way to our universe.   Another possibility is that distributions for the asymmetries strongly prefer $x$ near unity, so the helium abundance again becomes large.

To summarize: at $v/v_0 > 10^2$ weak interactions freeze out before the QCD phase transition leading to a helium abundance that depends on cosmic asymmetries.  The observed weak scale can be understood from the anthropic cost of helium domination provided such large $v$ universes are less probable than our own.  This can result if new physics below 10 TeV cuts off the fine-tuning in $v$ or if other environmental effects dominate at large $v$.

%%%%%%%%%%%%%%%%%%%%%%%%%%%
\section{Scanning Heavy Flavor Yukawa Couplings}
\label{sec:HF}
%%%%%%%%%%%%%%%%%%%%%%%%%%%%
In our universe, as the temperature drops below the mass of a heavy generation fermion, they rapidly decay and become cosmologically irrelevant.  However, since we scan Yukawa couplings of the first generation we should also consider scanning the Yukawa couplings of the heavy generations.  At low values of these Yukawa couplings could heavy flavors affect the helium abundance?  At $v/v_0> 10^2$, when the $u/d$ ratio is set before the QCD phase transition, this could occur both by heavy flavor contributions to the conditions for chemical equilibria that determine $u/d$ and by late decays of the heavy state affecting either $u/d$, $n/p$ or the helium abundance.   However, we ignore universes with very large $v/v_0$ since we have argued in the previous section that they must be less probable than our own for our understanding of the weak scale.  We discuss the relevant case of $v/v_0 < 10^2$.

The cosmic asymmetry in a heavy flavor allows a component that is unable to annihilate as the temperature drops below its mass.  If this component decays late, there are several ways it could affect the helium abundance.  Decays after 100s could lead to decay products which dissociate helium producing protons.  For the case of a muon, we find that the resulting electromagnetic shower only leads to significant dissociation at values of $v/v_0 \gg 10^2$ (unless $\eta_\mu \gg \eta_B$, which we would not expect).  Another possibility is that the heavy flavor decays produce charged pions between 1s and 100s and the reaction $\pi^+ n \rightarrow \pi^0 p$ (or $\pi^- p \rightarrow \pi^0 n$), so that a universe which would have had $n/p$ very close to unity ends up not being helium dominated.  For the muon we find that there is a very small region of parameter space where this could happen at  $10 < v/v_0 < 10^2$ and $y_\mu/ y_{\mu0} \sim 10^{-2}$.  We assume that this large reduction in the muon Yukawa coupling makes such universes less probable than our own.

Finally, it could be that heavy flavor decays directly produce protons, avoiding what would otherwise be helium domination.  This happens for a long-lived strange quark decaying via $\Lambda(uds) \rightarrow p \pi^-$. Again we find that there is a very small region of parameter space where this could happen, at  $10 < v/v_0 < 10^2$ and $y_s/ y_{s0} \sim 10^{-2}$, and assume such a small strange quark Yukawa coupling makes such universes less probable than our own.

%%%%%%%%%%%%%%%%%%%%%%%%%%%%%
\section{Conclusions}
\label{sec:concl}
%%%%%%%%%%%%%%%%%%%%%%%%%%%%%%%

In this paper we explored BBN in our neighborhood of the multiverse and found 
\begin{itemize}
\item  The observed value of the weak scale lies within the critical regime where BBN transitions from producing  all hydrogen to almost all helium, as shown in Figure~\ref{fig:BBNv}.  
An anthropic cost for processing hydrogen to helium determines the weak scale to be 
$v \sim (m_n - m_p)^{3/4} M_{pl}^{1/4}$ and no new physics is required to make the weak scale natural.  
\item The evidence that the first generation masses $m_u, m_d, m_e$ are determined anthropically from the existence of heavy elements and hydrogen is shown in Figures~\ref{fig:nuclearboundaries} and~\ref{fig:ude} and is significant. 
\item There is a finite volume in $(v, y_u, y_d, y_e)$ space where BBN does not produce too much helium, and where hydrogen and complex nuclei are stable, as shown in Figures~\ref{fig:BBNvmumd} and \ref{fig:BBNvmumdme}.  The observed values of these parameters is contained within, and lies near the boundary of, this anthropically allowed volume.  Therefore, the multiverse with scanning weak scale and first generation Yukawas can explain the observed values of these parameters.  This understanding of the weak scale and first generation masses can be preserved even when the Yukawa couplings of the heavy generations are allowed to scan.
\end{itemize}

We also examined large variations in the weak scale and found
\begin{itemize}
\item When more parameters are varied,  the weak scale can runaway to large values while maintaining primordial hydrogen.  We identified such runaways when the baryon to photon ratio or the Planck scale scan (Figure~\ref{fig:cosrun}), and also considered solutions to these runaways.
\item  For $v/v_0>10^2$ a new regime is entered where the helium abundance is determined by cosmic asymmetries and is independent of the weak scale.  We found that this does not invalidate the above results.  However, for them to be valid there must be some other reason why universes with these very large values of $v$ are disfavored.  Most exciting is the possibility that a Little Hierarchy, such as multi-TeV supersymmetry, makes these universes less probable. Another possibility is that anthropic arguments against such large values of $v$ may follow from either stellar burning or supernova explosions.  Finally, constraints on cosmic asymmetries might necessarily imply that $v/v_0>10^2$ leads to pure neutron universes, and an example of how this can arise was given.
\end{itemize}

\section*{Acknowledgments}

We are happy to thank Lance Dixon, Roni Harnik,  John March-Russell, Emanuele Mereghetti, Matt Reece, Jesse Thaler, Jay Wacker, Bob Wagoner, and Andre Walker-Loud for helpful conversations.  This work was supported in part by the Director, Office of Science, Office 
of High Energy and Nuclear Physics, of the US Department of Energy under 
Contract DE-AC02-05CH11231 and by the National Science Foundation under 
grants PHY-0457315, PHY-0855653 and PHY-1066293.  L.J.H. and J.T.R. thank the Aspen Center for Physics for hospitality, and L.J.H., D.P., and J.T.R. thank the Galileo Galilei Institute for Theoretical Physics for hospitality, while part of this work was completed.  J.T.R. was supported in part by the Miller Institute for Basic Research in Science.
\appendix

%%%%%%%%%%%%%%%%%%%%%%%%%%%
\section{Quark Mass Thresholds}
\label{app:LQCD}
%%%%%%%%%%%%%%%%%%%%%%%%%%%
In this appendix we discuss the implicit dependence of the QCD scale on the electroweak VEV due to quark mass thresholds.  Qualitatively, the physics can be described at one loop, allowing an unambiguous determination of $\Lambda_\text{QCD}$.  As $\alpha_s$ is run down from a high-scale boundary condition, heavy quarks should be integrated out at their respective mass scales.  The effective theory below the scale of a heavy quark has a different QCD beta function than the theory above that scale:
\be
\frac{d\alpha_s}{d\log \mu} = \left(11 - \frac{2}{3} n_f\right) \frac{\alpha_s^2}{2\pi} + ...
\ee
in which $n_f$ is the number of light quarks below the scale $\mu$.  At one loop, the beta function can be integrated, enforcing continuity of $\alpha_s$ at each quark mass threshold, and $\Lambda_\text{QCD}$ can be defined as the scale at which the QCD coupling blows up.  In the theory with three light quarks at the scale $\Lambda_\text{QCD}$, we have
\be
\Lambda_\text{QCD} \approx \mu^{7/9} \left(M_t M_b M_c\right)^{2/27} e^{-\frac{2\pi}{9 \alpha_s(\mu)}},
\ee
where $M_{t,b,c}$ are the pole masses of the top, bottom, and charm quarks, respectively, and $\mu$ is a scale large relative to the quark masses.  Then the QCD scale is approximately power-law dependent on $v$, with an exponent that depends on which Yukawa couplings are assumed to be fixed.

\begin{figure}[h!]
\begin{center} \includegraphics[width=1.00 \textwidth]{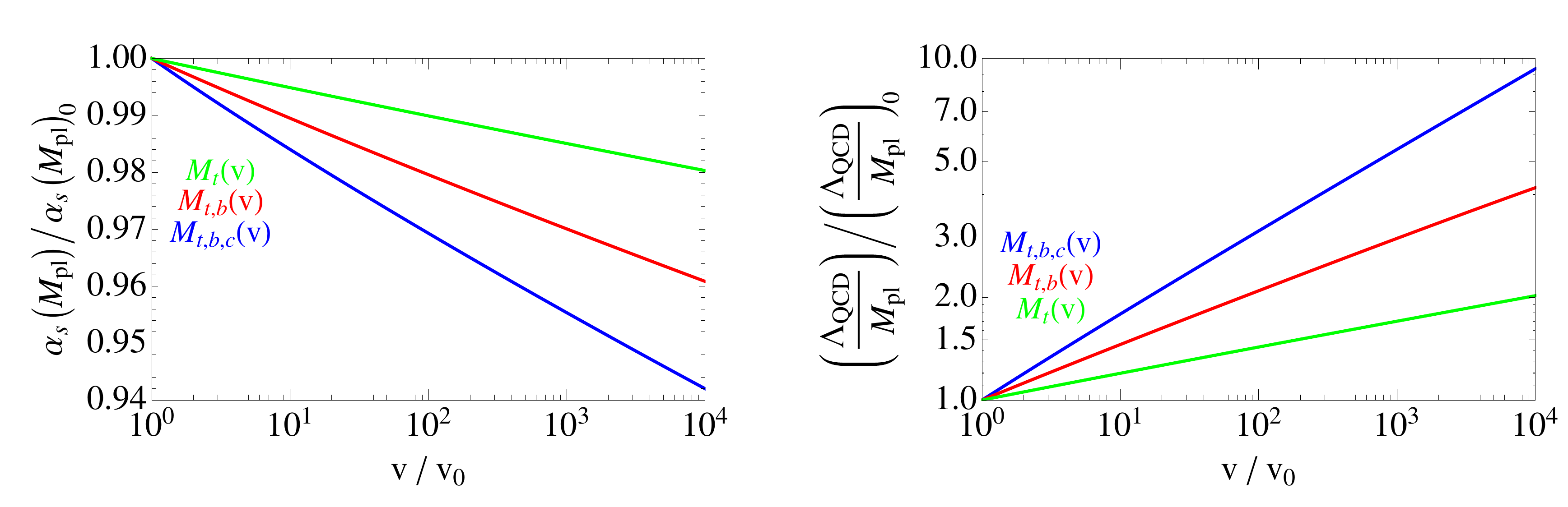}\end{center}
\caption{ \label{fig:alphas}
The left panel shows the high-scale value of $\alpha_s$ required to fix the ratio of $\Lambda_{\text{QCD}}/M_{\text{pl}}$ to its SM value as a function of $v$.  On the right, $\alpha_s\left(M_{\text{pl}}\right)$ is held fixed and the corresponding increase in $\Lambda_{\text{QCD}}/M_{\text{pl}}$ is shown as a function of $v$.  The different colors correspond to different assumptions about the $v$-dependence of the quark mass thresholds.  The top (top and bottom) [top, bottom, and charm] yukawas are held fixed along the green (red) [blue] line, so that the corresponding thresholds scale linearly with $v$.  All other quark masses are held fixed.  $\Lambda_{\text{QCD}}$ is calculated in the $\overline{\text{MS}}$ scheme at 2 loops in the three-flavor theory, with one loop matching at the mass thresholds.}
\end{figure}
	
The right-hand side of Figure~\ref{fig:alphas} shows the variation of $\Lambda_{\text{QCD}}$ with $v$, fixing the high-scale value of $\alpha_s$ to that of the SM\@.  The three different curves correspond to different assumptions about the heavy quark yukawas: for the blue curve, the top, bottom, and charm masses scale linearly with $v$; for the red, the top and bottom masses scale, while the charm mass is held fixed; and for the green, only the top mass scales, with the bottom and charm masses fixed.  In all cases, the three light quark masses are fixed to their values in our universe.  Here $\Lambda_{\text{QCD}}$ is defined in the theory with three light quarks at two loops in the $\overline{\text{MS}}$ scheme, with one loop matching at the heavy quark thresholds.  Because we have taken $\Lambda_{\text{QCD}}$ to be fixed in the rest of the paper, we have shown the resulting variation in terms of the ratio $\Lambda_{\text{QCD}}/M_{\text{pl}}$.  In terms of a fixed QCD scale, the quark mass thresholds have the effect of lowering the planck scale.  As described in Section~\ref{sec:scanetaMP}, this causes the reaction that produces deuterium to freeze out earlier, leading to reduced Helium production.

\begin{figure}[h!]
\begin{center} \includegraphics[width=.6 \textwidth]{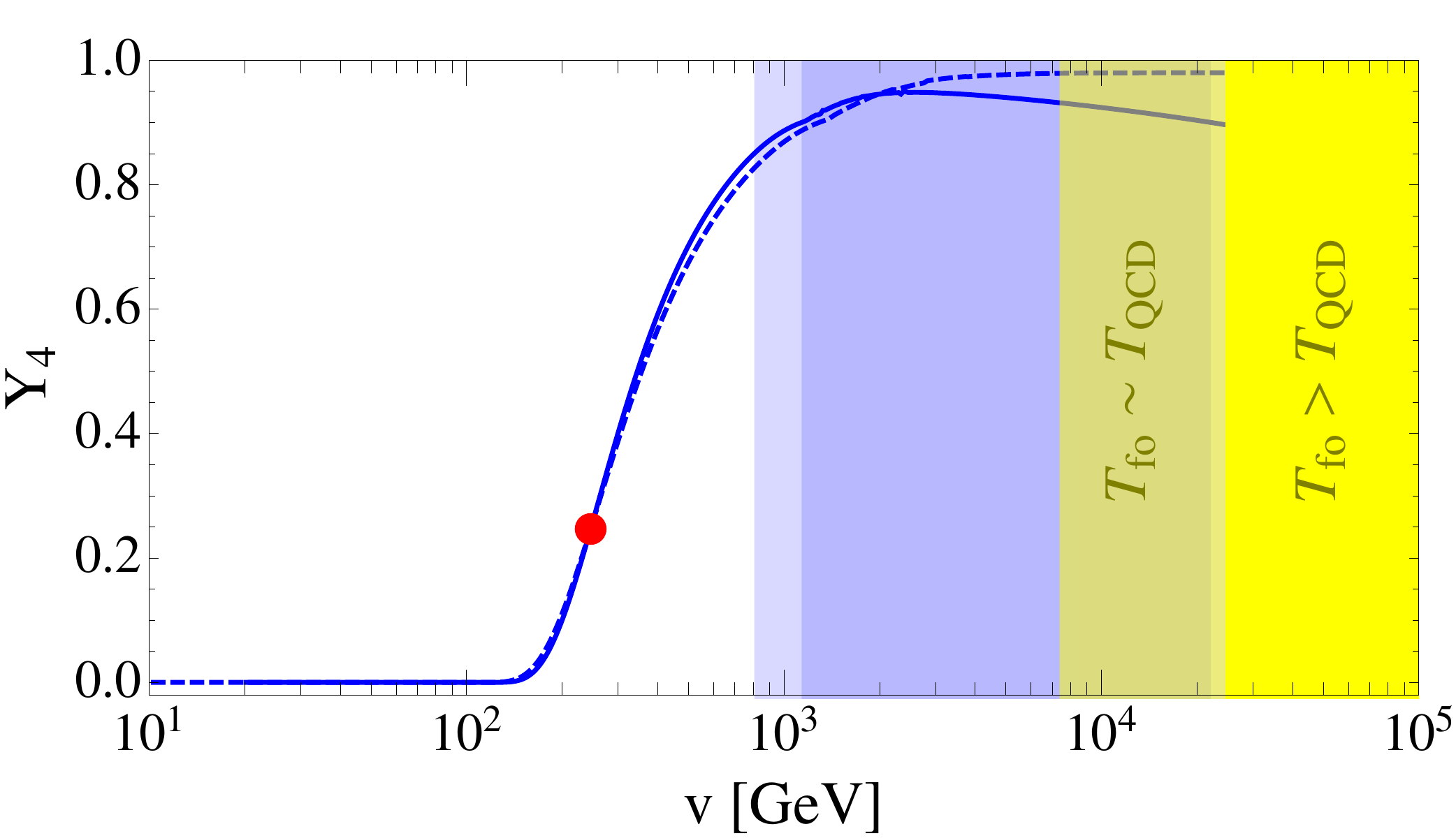}\end{center}
\caption{ \label{fig:BBNvFixas}
The analog of Figure~\ref{fig:BBNv} with the high-scale boundary condition for $\alpha_s$ held constant (solid curve).  The dashed curve is the same as Figure~\ref{fig:BBNv} and is the result of choosing $\alpha_s\left(M_{\text{pl}}\right)$ at each point such that the ratio $\Lambda_{\text{QCD}}/M_{\text{pl}}$ is constant with respect to $v$.  Light and dark shading correspond to 80\% and 90\% helium mass fractions, respectively.}
\end{figure}

In order to avoid this issue, we have taken the QCD scale (and thus the ratio $\Lambda_{\text{QCD}}/M_{\text{pl}}$) to be independent of $v$ throughout this work.  This is accomplished by fixing the second- and third-generation yukawas and performing a compensatory variation on the high-scale boundary condition for $\alpha_s$.  Because the QCD scale depends exponentially on $\alpha_s(M_{\text{pl}})$, the size of the required variation, shown in the left-hand side of Figure~\ref{fig:alphas}, is small even when $v$ is taken several orders of magnitude larger than its value in our universe.  Its size notwithstanding, this variation, taken by itself, spoils grand unification.  Restoring gauge coupling unification would induce a corresponding change in $\alpha_{\text{EM}}$ at low energies.  We have not required gauge coupling unification in this paper; however, the combination of GUTs with the implicit dependence of the gauge couplings on $v$ may motivate an expanded study that includes variations of $\alpha_\text{EM}$ in the future.


\begin{thebibliography}{0}
  
  \bibitem{Weinberg:1987dv}
S.~Weinberg,
%``Anthropic Bound on the Cosmological Constant,''
Phys.\ Rev.\ Lett.\  {\bf 59}, 2607 (1987).
%%CITATION = PRLTA,59,2607;%%

  \bibitem{Agrawal:1997gf}
~Agrawal, S.~M.~Barr, J.~F.~Donoghue and D.~Seckel,
%``The anthropic principle and the mass scale of the standard model,''
Phys.\ Rev.\  D {\bf 57}, 5480 (1998)
[arXiv:hep-ph/9707380].
%%CITATION = PHRVA,D57,5480;%%

%\cite{Martel:1997vi}
\bibitem{Martel:1997vi} 
  H.~Martel, P.~R.~Shapiro and S.~Weinberg,
  %``Likely values of the cosmological constant,''
  Astrophys.\ J.\  {\bf 492}, 29 (1998)
  [astro-ph/9701099].
  
%\cite{Damour:2007uv}  
\bibitem{Damour:2007uv}
T.~Damour and J.~F.~Donoghue,
%``Constraints on the variability of quark masses from nuclear binding,''
Phys.\ Rev.\  D {\bf 78}, 014014 (2008)
[arXiv:0712.2968 [hep-ph]].
%%CITATION = PHRVA,D78,014014;%%

 %\cite{ArkaniHamed:2005yv}
\bibitem{ArkaniHamed:2005yv} 
  N.~Arkani-Hamed, S.~Dimopoulos and S.~Kachru,
  %``Predictive landscapes and new physics at a TeV,''
  hep-th/0501082.
  %%CITATION = HEP-TH/0501082;%%
 
 %\cite{Froggatt:1978nt}
\bibitem{Froggatt:1978nt} 
  C.~D.~Froggatt and H.~B.~Nielsen,
  %``Hierarchy of Quark Masses, Cabibbo Angles and CP Violation,''
  Nucl.\ Phys.\ B {\bf 147}, 277 (1979). 
   
   %\cite{ArkaniHamed:1999dc}
\bibitem{ArkaniHamed:1999dc} 
  N.~Arkani-Hamed and M.~Schmaltz,
  %``Hierarchies without symmetries from extra dimensions,''
  Phys.\ Rev.\ D {\bf 61}, 033005 (2000)
  [hep-ph/9903417].

%\cite{Tegmark:2005dy}
\bibitem{Tegmark:2005dy} 
  M.~Tegmark, A.~Aguirre, M.~Rees and F.~Wilczek,
  %``Dimensionless constants, cosmology and other dark matters,''
  Phys.\ Rev.\ D {\bf 73}, 023505 (2006)
  [astro-ph/0511774].
  
  %\cite{Bousso:2007kq}
\bibitem{Bousso:2007kq} 
  R.~Bousso, R.~Harnik, G.~D.~Kribs and G.~Perez,
  %``Predicting the Cosmological Constant from the Causal Entropic Principle,''
  Phys.\ Rev.\ D {\bf 76}, 043513 (2007)
  [hep-th/0702115 [HEP-TH]].
  
%\cite{Carr:1979sg}
\bibitem{Carr:1979sg} 
  B.~J.~Carr and M.~J.~Rees,
  %``The anthropic principle and the structure of the physical world,''
  Nature {\bf 278}, 605 (1979).
  
  %\cite{Janka:2012wk}
\bibitem{Janka:2012wk} 
  H.~-T.~Janka,
  %``Explosion Mechanisms of Core-Collapse Supernovae,''
  Ann.\ Rev.\ Nucl.\ Part.\ Sci.\  {\bf 62}, 407 (2012)
  [arXiv:1206.2503 [astro-ph.SR]].
  %%CITATION = ARXIV:1206.2503;%%
  
   %\cite{Harnik:2006vj}
\bibitem{Harnik:2006vj} 
  R.~Harnik, G.~D.~Kribs and G.~Perez,
  %``A Universe without weak interactions,''
  Phys.\ Rev.\ D {\bf 74}, 035006 (2006)
  [hep-ph/0604027].
  %%CITATION = HEP-PH/0604027;%%

 %\cite{Graesser:2006ft}
\bibitem{Graesser:2006ft} 
  M.~L.~Graesser and M.~P.~Salem,
  %``The scale of gravity and the cosmological constant within a landscape,''
  Phys.\ Rev.\ D {\bf 76}, 043506 (2007)
  [astro-ph/0611694].
  %%CITATION = ASTRO-PH/0611694;%%
  
  %\cite{Walker-Loud:2014iea}
\bibitem{Walker-Loud:2014iea} 
  A.~Walker-Loud,
  %``Nuclear Physics Review,''
  PoS LATTICE {\bf 2013}, 013 (2014)
  [arXiv:1401.8259 [hep-lat]].
  %%CITATION = ARXIV:1401.8259;%%


  
%\cite{Hall:2007ja}
\bibitem{Hall:2007ja} 
  L.~J.~Hall and Y.~Nomura,
  %``Evidence for the Multiverse in the Standard Model and Beyond,''
  Phys.\ Rev.\ D {\bf 78}, 035001 (2008)
  [arXiv:0712.2454 [hep-ph]].
  
    %\cite{Bedaque:2010hr}
\bibitem{Bedaque:2010hr} 
  P.~F.~Bedaque, T.~Luu and L.~Platter,
  %``Quark mass variation constraints from Big Bang nucleosynthesis,''
  Phys.\ Rev.\ C {\bf 83}, 045803 (2011)
  [arXiv:1012.3840 [nucl-th]].
  %%CITATION = ARXIV:1012.3840;%%
  %17 citations counted in INSPIRE as of 16 Jan 2014



  
  %\cite{Arbey:2011nf}
\bibitem{Arbey:2011nf} 
  A.~Arbey,
  %``AlterBBN: A program for calculating the BBN abundances of the elements in alternative cosmologies,''
  Comput.\ Phys.\ Commun.\  {\bf 183}, 1822 (2012)
  [arXiv:1106.1363 [astro-ph.CO]].
  %%CITATION = ARXIV:1106.1363;%%
  %7 citations counted in INSPIRE as of 16 Jan 2014
  
  %\cite{Berengut:2013nh}
\bibitem{Berengut:2013nh} 
  J.~C.~Berengut, E.~Epelbaum, V.~V.~Flambaum, C.~Hanhart, U.~-G.~Meissner, J.~Nebreda and J.~R.~Pelaez,
  %``Varying the light quark mass: impact on the nuclear force and Big Bang nucleosynthesis,''
  Phys.\ Rev.\ D {\bf 87}, no. 8, 085018 (2013)
  [arXiv:1301.1738 [nucl-th]].
  %%CITATION = ARXIV:1301.1738;%%
  %15 citations counted in INSPIRE as of 16 Jan 2014
  
  %\cite{Rupak:1999rk}
\bibitem{Rupak:1999rk} 
  G.~Rupak,
  %``Precision calculation of n p ---> d gamma cross-section for big bang nucleosynthesis,''
  Nucl.\ Phys.\ A {\bf 678}, 405 (2000)
  [nucl-th/9911018].
  %%CITATION = NUCL-TH/9911018;%%
  %67 citations counted in INSPIRE as of 16 Jan 2014

  %\cite{Beane:2002vs}
\bibitem{Beane:2002vs} 
  S.~R.~Beane and M.~J.~Savage,
  %``Variation of fundamental couplings and nuclear forces,''
  Nucl.\ Phys.\ A {\bf 713}, 148 (2003)
  [hep-ph/0206113].
  %%CITATION = HEP-PH/0206113;%%
  %96 citations counted in INSPIRE as of 29 Jan 2014

  %\cite{MacDonald:2009vk}
\bibitem{MacDonald:2009vk} 
  J.~MacDonald and D.~J.~Mullan,
  %``Big Bang Nucleosynthesis: The Strong Force meets the Weak Anthropic Principle,''
  Phys.\ Rev.\ D {\bf 80}, 043507 (2009)
  [arXiv:0904.1807 [astro-ph.CO]].
  %%CITATION = ARXIV:0904.1807;%%
  %8 citations counted in INSPIRE as of 16 Jan 2014
  
  %\cite{Bradford:2009}
\bibitem{Bradford:2009}
  R.A.W.~Bradford,
  J.\ Astrophys.\ Astr. {\bf 30}, 119 (2009).
   
  %\cite{Katz:1995up}
\bibitem{Katz:1995up} 
  N.~Katz, D.~H.~Weinberg and L.~Hernquist,
  %``Cosmological simulations with TreeSPH,''
  Astrophys.\ J.\ Suppl.\  {\bf 105}, 19 (1996)
  [astro-ph/9509107].
  %%CITATION = ASTRO-PH/9509107;%%
  
  %\cite{Bousso:2009ks}
\bibitem{Bousso:2009ks} 
  R.~Bousso, L.~J.~Hall and Y.~Nomura,
  %``Multiverse Understanding of Cosmological Coincidences,''
  Phys.\ Rev.\ D {\bf 80}, 063510 (2009)
  [arXiv:0902.2263 [hep-th]].
  %%CITATION = ARXIV:0902.2263;%%
  
  %\cite{Bousso:2013rda}
\bibitem{Bousso:2013rda} 
  R.~Bousso and L.~Hall,
  %``Why Comparable? A Multiverse Explanation of the Dark Matter-Baryon Coincidence,''
  Phys.\ Rev.\ D {\bf 88}, 063503 (2013)
  [arXiv:1304.6407 [hep-th]].
  
%\cite{Bhattacharya:2014ara}
\bibitem{Bhattacharya:2014ara} 
  T.~Bhattacharya, M.~I.~Buchoff, N.~H.~Christ, H.~-T.~Ding, R.~Gupta, C.~Jung, F.~Karsch and Z.~Lin {\it et al.},
  %``The QCD phase transition with physical-mass, chiral quarks,''
  arXiv:1402.5175 [hep-lat].
   
  \end{thebibliography}
\end{document}